\documentclass[prd,twocolumn,nofootinbib,superscriptaddress]{revtex4-2}

\usepackage{comment} 

\usepackage{graphicx}
\usepackage{amsmath,bm,amssymb,amsfonts,dsfont}
\usepackage[usenames,dvipsnames]{xcolor}
\usepackage[normalem]{ulem}
\usepackage{url}
\usepackage{array}
\usepackage{booktabs}
\usepackage{multirow}
\usepackage{float}
\usepackage[colorlinks  = true,
            linkcolor   = NavyBlue,
            urlcolor    = NavyBlue,
            citecolor   = NavyBlue,
            anchorcolor = NavyBlue]{hyperref}
\usepackage{cprotect}

\usepackage{mathtools,slashed}

\usepackage{appendix}

\usepackage{bbold}

\usepackage[compat=1.1.0]{tikz-feynman} 
\tikzfeynmanset{every vertex/.style={/tikz/font=\small}}

\def \aap  {A\&A}

\def \apj  {ApJ}

\def \mnras {MNRAS}

\begin{document}

\title{A Precise Measurement of the {\it Fermi}-LAT Galactic Center Excess Morphology and Spectrum}

\author{Mattia Di Mauro}\email{dimauro.mattia@gmail.com}
\affiliation{Istituto Nazionale di Fisica Nucleare, Sezione di Torino, Via P. Giuria 1, 10125 Torino, Italy}

\date{\today}

\begin{abstract}
We present a new \emph{Fermi}-LAT analysis of the Galactic-center excess (GCE) designed to substantially reduce the dominant systematic uncertainties associated with interstellar-emission and source modeling in the inner Galaxy.
Using an optimized multi-step fitting procedure together with an iterative source-finding pipeline, we achieve a markedly improved agreement between data and model, reducing fractional residuals to $\lesssim 10\%$ over a $40^\circ\times 40^\circ$ region centered on the Galactic center.
We analyze a suite of \texttt{GALPROP}-based interstellar-emission models (IEMs) and complementary analysis variants (Galactic-plane masking, fits restricted to $1$--$10$~GeV, and weighted-likelihood fits) to quantify robustness.
The reconstructed surface-brightness profile is strongly centrally concentrated and is well described by an approximately spherical generalized Navarro-Frenk-White morphology with inner slope $\gamma \simeq 1.15$.
Bulge-tracing templates (nuclear bulge plus boxy bulge) fail to reproduce the full radial morphology, most notably for line-of-sight angles around $\theta\simeq 1^\circ$--$2^\circ$ and at $\theta \gtrsim 8^\circ$, whereas the DM-motivated component provides a good description over the full angular range.
Moreover, the DM component remains highly significant across all IEMs and analysis choices, including fits that simultaneously include the bulge templates.
We also provide an updated measurement of the GCE spectrum from $0.5$ to $1000$~GeV, confirming a peak at a few GeV and setting stringent constraints above tens of GeV, where we obtain only upper limits at the level $E^2\Phi \lesssim 10^{-8}$~GeV\,cm$^{-2}$\,s$^{-1}$\,sr$^{-1}$.
These results deliver a sharpened and systematically controlled characterization of the GCE morphology and spectrum, enabling more incisive tests of astrophysical and dark-matter interpretations.
\end{abstract}

\maketitle

\emph{Introduction.—}
A wealth of gravitational evidence establishes the presence of dark matter (DM) from galactic to cosmological scales, yet its microscopic identity remains unknown~\cite{Bertone:2010zza,Bertone:2016nfn,Cirelli:2024ssz}.
Any particle DM explanation requires physics beyond the Standard Model (SM) and must reproduce the measured relic density $\Omega_{\rm DM}h^2\simeq0.12$~\cite{Aghanim:2018eyx}.
A widely studied benchmark is thermal weakly-interactive-massive particles (WIMPs), for which freeze-out can yield the observed abundance for $m_{\rm DM}\sim10$--$10^3$~GeV and electroweak-scale interactions.
The WIMP scenario is probed by direct detection, collider searches, and indirect searches for annihilation/decay products such as $\gamma$ rays, antimatter, and neutrinos~\cite{Schumann:2019eaa,Boveia:2018yeb,Gaskins:2016cha}.
$\gamma$ rays are particularly powerful in indirect searches because they propagate undeflected and preserve spectral and morphological information.

The Galactic Center (GC), expected to host the largest DM density in the local Universe~\cite{Pieri:2009je}, is therefore a prime indirect-detection target.
Multiple analyses of \emph{Fermi}-LAT data have reported an excess of GeV $\gamma$ rays over standard diffuse-emission expectations, the Galactic Center Excess (GCE)~\cite{Goodenough:2009gk,Hooper:2010mq,Boyarsky:2010dr,Hooper:2011ti,Abazajian:2012pn,Gordon:2013vta,Abazajian:2014fta,Daylan:2014rsa,Calore:2014nla,Calore:2014xka,TheFermi-LAT:2015kwa,TheFermi-LAT:2017vmf,DiMauro:2019frs,DiMauro:2021raz,Cholis:2021rpp}.
Template-based studies find a spectrum peaking at a few GeV with emission extending to $\sim\!50$~GeV and a morphology well described by a generalized Navarro-Frenk-White (NFW)-like profile with inner slope $\gamma\simeq 1.2$ (see, e.g., \cite{Calore:2014nla,DiMauro:2021raz,Cholis:2021rpp}).
These characteristics are consistent with an annihilating WIMP interpretation, e.g.\ $m_{\rm DM}\!\sim\!30$--$60$~GeV into hadronic~\cite{Calore:2014xka,DiMauro:2021qcf,Kong:2025ccv} or leptonic~\cite{Koechler:2025ryv,Kong:2025ccv} final states, with an annihilation rate comparable to the thermal benchmark.

Astrophysical scenarios remain a viable alternative, most notably an unresolved population of millisecond pulsars (MSPs) in the Galactic bulge~\cite{Bartels:2015aea,Lee:2015fea,Macias:2016nev,Bartels:2017vsx,Manconi:2024tgh}.
Analyses employing stellar-bulge templates (nuclear plus boxy/peanut bulge) as MSP tracers~\cite{Gordon:2013vta,Macias:2016nev,Coleman:2019kax,Abazajian:2020tww,Calore:2021jvg,Pohl:2022nnd,Manconi:2024tgh} can outperform purely spherical DM-motivated templates, suggesting departures from strict spherical symmetry for the GCE morphology.
In particular, Refs.~\cite{Macias:2016nev,Coleman:2019kax,Abazajian:2020tww,Calore:2021jvg,Manconi:2024tgh} argued that including nuclear-bulge (NB) and boxy-bulge (BB) components renders any additional DM-like contribution compatible with zero flux and significance, implying strong constraints on a putative DM signal~\cite{Abazajian:2020tww,Manconi:2025ogr}.

From the particle-physics perspective, a DM interpretation is further constrained by null results from other searches. Current direct- and indirect-detection limits at weak-scale masses exclude large regions of parameter space in scenarios where the same portal controls both freeze-out and elastic scattering~\cite{Arcadi:2017kky,Arcadi:2019lka,DiMauro:2023tho,Arcadi:2024ukq,DiMauro:2025jia,Kong:2025ccv}.
This motivates resonant-funnel realizations with $m_{\rm DM}\simeq m_{\rm med}/2$~\cite{DiMauro:2023tho,Koechler:2025ryv,DiMauro:2025jia,Kong:2025ccv}, as well as more general secluded scenarios where the relic density is set by annihilations into dark-sector mediators with small portal couplings~\cite{Pospelov:2007mp,Pospelov:2008jd,DiMauro:2025jsb,DiMauro:2025uxt}.

\begin{figure}
  \centering
  \includegraphics[width=0.99\linewidth]{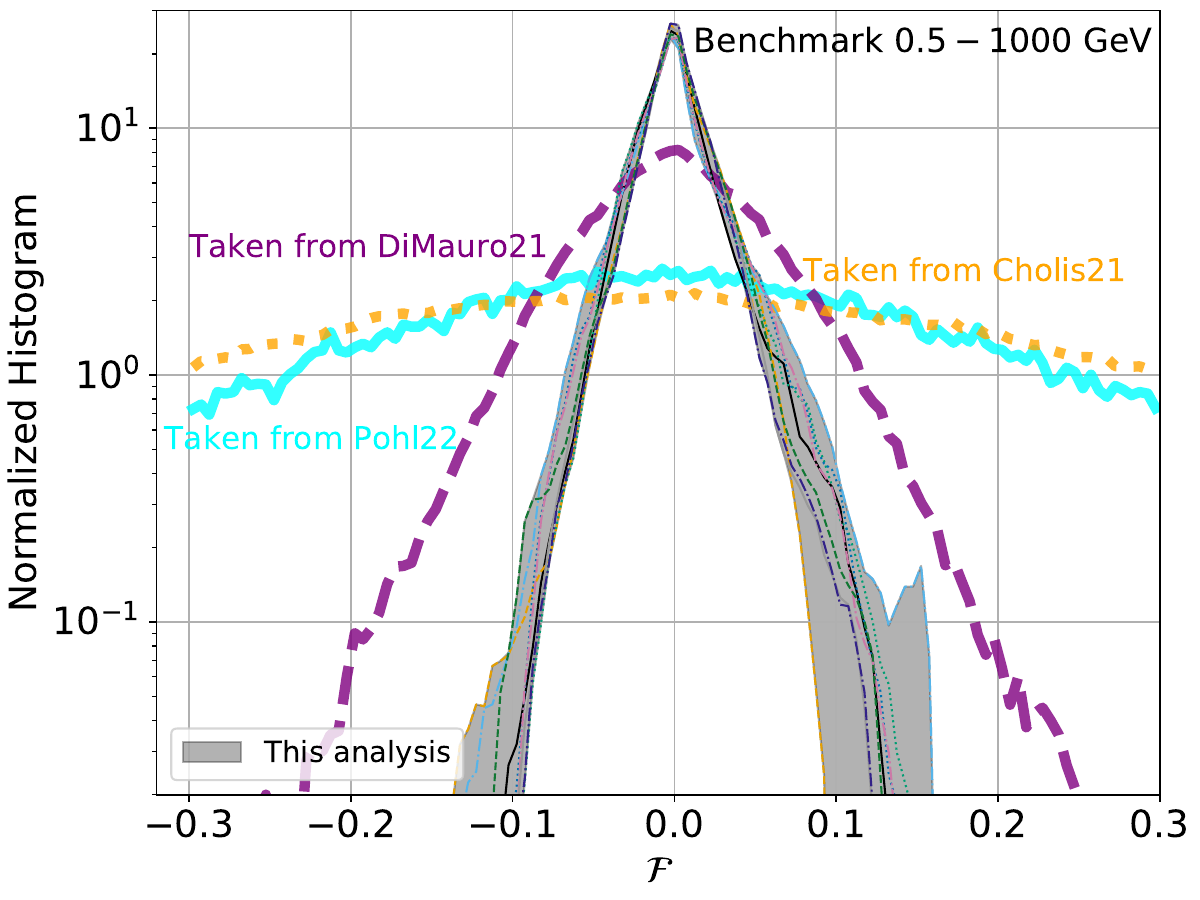}
  \caption{Normalized histogram of the fractional residuals $\mathcal{F}_i$ (see Eq.~\ref{eq:F}). We show the results reported in \cite{DiMauro:2021prd} (DiMauro21, purple band), \cite{Pohl:2022nnd} (Pohl22, cyan band), and \cite{Cholis:2021rpp} (Cholis21, orange dotted). We also display the outcome of our analysis for several IEMs, including those adopted in the aforementioned works (colored curves and gray band).}
\label{fig:residuals}
\end{figure}

Independently of the underlying interpretation, significant progress in a precise derivation of the GCE characteristics is limited because in the GC region the observed $\gamma$-ray intensity is dominated by Galactic interstellar emission (IE).
In practice, the precision with which the GCE spectrum and morphology can be determined is limited by systematics in the interstellar emission model (IEM), degeneracies among its components, and the modeling of the source population (catalog and newly detected sources)~\cite{Leane:2019uhc,Chang:2019ars,Zhong:2019ycb,Calore:2021jvg,List:2025qbx}.
Most previous studies (see, e.g., \cite{Gordon:2013vta,Macias:2016nev,Coleman:2019kax,Abazajian:2020tww,DiMauro:2021prd,Cholis:2021rpp}) rely on IEM realizations that leave fractional residuals, evaluated as
\begin{equation}
\label{eq:F}
\mathcal{F}_i = 2\cdot \frac{\mathcal{C}_{M,i}-\mathcal{C}_{D,i}}{\mathcal{C}_{M,i}+\mathcal{C}_{D,i}},
\end{equation}
where $\mathcal{C}_{M,i}$ ($\mathcal{C}_{D,i}$) is the number of model (data) counts in the $i$-th pixel, and which can reach the $\sim 30$--$40\%$ level in the inner Galaxy.
This is illustrated in Fig.~\ref{fig:residuals}, which shows the normalized histogram of $\mathcal{F}_i$ using the residual distributions reported in \cite{DiMauro:2021prd} (DiMauro21, $E>300$~MeV), \cite{Pohl:2022nnd} (Pohl22, $E>600$~MeV), and \cite{Cholis:2021rpp} (Cholis21, $E>500$~MeV).

In this Letter we present a new analysis of \emph{Fermi}-LAT GC data designed to substantially mitigate these modeling uncertainties and thereby enable a more robust determination of the GCE properties.
With our pipeline, using data between $0.5$--$1000$~GeV, and different IEMs that include the ones from \cite{DiMauro:2021prd,Cholis:2021rpp,Pohl:2022nnd}, the fractional residuals reach values, for most of the GC Region-of-Interest (ROI), that are at $\lesssim 10\%$ (Fig.~\ref{fig:residuals}), i.e.~significantly smaller than in previous analyses.
By improving the treatment of diffuse emission, $\gamma$-ray sources, and their associated degeneracies, we obtain more precise measurements of the morphology and spectral energy distribution (SED) of the GCE, providing a sharpened empirical target for both particle-DM and astrophysical interpretations.

\emph{Analysis technique.—}
In our benchmark analysis, we use $\sim$16~years of \emph{Fermi}-LAT \cite{2018arXiv181011394B} observations in a $40^\circ\times40^\circ$ ROI centered on the GC.
We select \texttt{P8R3\_SOURCEVETO} events and apply standard data-quality selections \cite{2018arXiv181011394B}.
To minimize the impact of IEM systematics, which are most pronounced below 1~GeV, we use $1$--$10$~GeV when studying the spatial morphology of the GCE. The choice of selecting 10 GeV as an upper bound is due to the fact that at very-high-energy the GCE flux is very small and thus restricting the energy range makes the fit of IEM and source SED less complicated to fit.
For the SED evaluation, we instead consider $0.5$--$1000$~GeV photons in order to characterize the low- and high-energy tail in addition to the peak at a few GeV.
The data are binned with 8 energy bins per decade and a spatial pixel size of $0.08^\circ$.
We perform a binned-likelihood analysis with \texttt{Fermipy}, which provides a high-level interface to the \texttt{Fermitools} for standard \emph{Fermi}-LAT likelihood workflows~\cite{2017ICRC...35..824W}.
We refer to this data selection and analysis configuration as the \textbf{Benchmark} setup.

The central rationale of our approach is to mitigate the dominant systematics that arise when modeling lines of sight toward the GC, while maintaining numerical stability in a source-crowded ROI.
To this end, we adopt a staged fitting strategy:
(i) we initialize the model with the 4FGL-DR4 source catalog~\cite{Fermi-LAT:2022byn,Ballet:2023qzs} and the IE components (see the next part for their description), and obtain a first background model via iterative optimization of the spectral parameters;
(ii) we refine the model with a likelihood maximization in which only the most degenerate components (diffuse templates and the brightest/innermost sources) are left free to vary; and
(iii) we iteratively search for additional point-like and extended source candidates using a sequence of test-statistic (TS) thresholds, followed by a final re-optimization and refit of the spectral shapes.
This procedure yields residual maps that are largely featureless at the $\lesssim 10\%$ level (Fig.~\ref{fig:residuals}) and enables a model-agnostic and precise extraction of the GCE properties.
Full details of the data selection and analysis pipeline are provided in Appendix~\ref{app:analysis}.

Because the GC region is dominated by bright, structured diffuse emission, imperfect modeling of the Galactic plane can bias residuals and inflate TS, especially at low energies.
Therefore, in addition to the \textbf{Benchmark} setup, we consider two complementary analysis choices designed to reduce the impact of diffuse mismodeling; for both, we select \texttt{SOURCEVETO} events in the range $0.5$--$1000$~GeV:
\begin{itemize}
  \item \textbf{Weighted-likelihood analysis.}
  We employ the weighted-likelihood method used in \emph{Fermi}-LAT analyses to reduce sensitivity to diffuse-model systematics, most notably along the Galactic plane and below $\sim$1~GeV~\cite{Bruel:2021mrd,Fermi-LAT:2019yla}.
  In this approach, each spatial/energy bin is assigned a weight $\epsilon\in[0,1]$ derived from the model-predicted foreground intensity, so that bins with large expected backgrounds are down-weighted in the likelihood and do not dominate the fit. Typically, only pixels within a few degrees from the Galactic plane and for energies lower than 1 GeV have $\epsilon \ll 1$.

  \item \textbf{Galactic-plane mask.}
  As a complementary (and more conservative) approach, we repeat the analysis after masking the Galactic plane region $|b|<2^\circ$.
  This removes the brightest diffuse emission and many of the regions where gas-template imperfections and source distribution systematics are largest, at the cost of reduced statistics and diminished sensitivity to the GCE.
\end{itemize}

\emph{Interstellar-emission models.—}
Diffuse Galactic emission dominates the $\gamma$-ray data toward the inner Galaxy ($\sim 90\%$ of the counts) and is the leading source of systematic uncertainty in GCE measurements, while the GCE contributes only $\sim 5$--$10\%$.
We therefore repeat the analysis with nine IEMs based on \texttt{GALPROP}~\cite{Porter:2021tlr}, which predicts diffuse $\gamma$ rays by solving the cosmic-ray (CR) transport equation for specified CR source, propagation, gas, and interstellar radiation field (ISRF) inputs.
All IEMs include inverse Compton (IC), bremsstrahlung, and $\pi^0$-decay emission, together with large-scale templates such as the \emph{Fermi} bubbles and Loop~I.
All the IEMs considered here are derived within the \texttt{GALPROP} framework because it provides a physically grounded and self-consistent prediction of the different diffuse $\gamma$-ray emission components by solving the CR transport equation given explicit assumptions on sources, propagation, gas, and the interstellar radiation field.
Some of the models further decompose gas-correlated and/or IC emission into Galactocentric rings, allowing the fit to float the relative normalization of different line-of-sight regions.
We refer to the original references and to Appendix~\ref{app:IEMs} for full technical details, and summarize here the main differences among the adopted models.

\begin{itemize}
  \item \textbf{DiMauro21}~\cite{DiMauro:2021prd}: adopts the Yusifov pulsar source distribution~\cite{Yusifov:2004fr} and implements a compact 7-template set (two gas-correlated components, three IC subcomponents, and inner/outer bubble templates).
  \item \textbf{Cholis21}~\cite{Cholis:2021rpp}: tuned to local primary and secondary CR data; same compact structure as \textbf{DiMauro21}. We use the three best models in Ref.~\cite{Cholis:2021rpp} (\texttt{8l, ch, bf}).
  \item \textbf{Galp21}~\cite{Porter:2021tlr}: public \texttt{GALPROP} release models using the ISRF of Ref.~\cite{2012A&A...545A..39R}. The diffuse components are provided in multiple Galactocentric rings (yielding $\sim$20 templates in our implementation). We consider three models labeled \texttt{SA0}, \texttt{SA50}, and \texttt{SA100}, which assume different CR source distributions.
  \item \textbf{Macias19}~\cite{Coleman:2019kax}: ring-decomposed IEM with updated dust-based gas corrections and explicit dust-residual templates; includes 6 IC rings and 4 gas rings (19 templates in our implementation) and is often paired with NB and BB templates.
  \item \textbf{Pohl22}~\cite{Pohl:2022nnd}: update of \textbf{Macias19} with a revised reconstruction of inner-Galaxy H\,\textsc{i}, leaving the ring/template structure essentially unchanged.
\end{itemize}


\emph{Goodness of fit for DM and boxy-bulge templates.—}
Within our benchmark setup, the IEMs that provide the best overall description of the \emph{Fermi}-LAT data are the \texttt{Galp21} realizations, with \texttt{Galp21\_SA100} yielding the largest likelihood.
Relative to this best-fitting reference, the remaining models are disfavored by
$\Delta\log\mathcal{L}_{\rm DM}\equiv \log\mathcal{L}_{\rm DM}-\log\mathcal{L}_{\rm DM}^{\max}$ values at the level of $\mathcal{O} (10^3)$ (see Tab.~\ref{tab:E_summary}).
Notably, the \texttt{Macias19} and \texttt{Pohl22} IEMs---often argued to outperform simpler baselines in GC analyses~\cite{Coleman:2019kax,Abazajian:2020tww}---yield likelihoods comparable to the much simpler \texttt{DiMauro21} and \texttt{Cholis21} models once the same data selection and fitting procedure are applied.
We use Tab.~\ref{tab:E_summary} to quantify, in a controlled way, how the inferred GCE properties and interpretation depend on the IEM choice.

For each IEM we detect the DM-motivated template (a gNFW morphology with inner slope $\gamma\simeq 1.1$--$1.2$) with high statistical significance.
The corresponding test statistic spans $\mathrm{TS}_{\rm DM}\simeq (0.6$--$2.2)\times 10^4$ across the model set, with the lowest value obtained for \texttt{Pohl22} and the largest values in the \texttt{Galp21} family.
Under the usual asymptotic mapping, these TS values correspond to nominal Gaussian significances of order $\sqrt{\mathrm{TS}}$, i.e.\ $\sim 80$--$150\sigma$.\footnote{For nested hypotheses, and when the conditions of Chernoff's theorem are satisfied, the detection significance is often quoted as $\simeq \sqrt{\mathrm{TS}}$.}

We next assess how the fit changes when the DM template is replaced by the bulge-tracing components adopted in Ref.~\cite{Coleman:2019kax} (BB plus NB).
Tab.~\ref{tab:E_summary} reports $\Delta\log\mathcal{L}_{\rm NB}$, defined as the change in log-likelihood when fitting with NB+BB in place of the DM template (all other ingredients treated consistently).
For all IEMs except \texttt{Pohl22}, $\Delta\log\mathcal{L}_{\rm NB}<0$, implying that NB+BB alone provides a worse description of the data than the DM template; for the \texttt{Galp21} models the degradation is typically at the level of $\mathcal{O}(10^2)$ in $\Delta\log\mathcal{L}$.
Nevertheless, the bulge templates are themselves detected with large TS values (TS$_{\rm B/N}$ in Tab.~\ref{tab:E_summary}), reflecting that the inner Galaxy can contain substantial emission with morphologies correlated with the stellar bulge, even when such templates do not fully account for the GCE.

Finally, we consider a combined model including both the DM template and the NB/BB components.
In all cases, adding NB+BB on top of the DM-only baseline improves the fit only minimally, with $\Delta\log\mathcal{L}_{\rm DNB}>0$ between a few tens up to a few hundreds, with the IEM \texttt{Pohl22} that provides the larger improvement with 347.
This indicates that for most of the models addition of bulge-correlated structures does not appear to be essential for capturing residual emission beyond the DM-only model.
Crucially, the DM component remains highly significant in the joint fit: TS$_{\rm DM}^{\rm DNB}$ lies in the range $8\times 10^2$--$1.5\times 10^4$ across the IEM set, implying a robust detection even in the presence of stellar-bulge tracers.\footnote{The corresponding nominal Gaussian significance is $\sim 28$--$120\sigma$ under the usual $\sqrt{\mathrm{TS}}$ mapping.}
This behavior differs from the conclusions of Refs.~\cite{Macias:2016nev,Coleman:2019kax,Abazajian:2020tww}; we note, however, that those analyses typically exhibited substantially larger diffuse-model residuals, whereas our pipeline achieves residuals at the $\lesssim 10\%$ level over the ROI (Fig.~\ref{fig:residuals}).

When fitting with DM, NB, and BB simultaneously, the statistical weight carried by NB and BB in the joint fit is more model dependent: in several IEMs the bulge TS values are substantially reduced relative to the NB+BB-only fit, consistent with strong degeneracies among extended templates in the inner Galaxy.
An important exception is \texttt{Pohl22}, where NB and BB retain very large TS values and the DM TS is comparatively reduced.

Finally, we consider a model including only the DM and NB templates (i.e.\ without BB). 
This yields only a modest improvement relative to the DM-only fit, and the NB component is detected with TS ranging from $\sim 50$ up to the maximum value obtained for \texttt{Pohl22}, TS$=2763$.

The results and main outcome of the analysis discussed here are very similar when using \textbf{Benchmark} setup with $1-10$ GeV and $0.5-1000$ GeV energy selection and the \textbf{Weighted-likelihood} and \textbf{Galactic-plane mask} setups. In Appendix \ref{app:results} we report the results obtained with the different cases and IEMs.

Overall, within our refined analysis framework we find that a DM-like excess component persists with very high significance even when including physically motivated bulge templates intended to trace an unresolved MSP population~\cite{Coleman:2019kax}.
At the same time, the preference of the DM template over the bulge ones is not statistically significant enough to establish a DM origin of the GCE, and remains sensitive to the adopted IEM.
This motivates the systematic treatment of diffuse-emission uncertainties adopted in the remainder of this work.

\begin{table*}[t]
\centering
\begin{tabular}{lrrrrrrrrrr}
\hline
Model & $\Delta\log\mathcal{L}_{\rm DM}$ & TS$_{\rm DM}$ & $\Delta\log\mathcal{L}_{\rm NB}$ & TS$_{\rm B/N}$ & $\Delta\log\mathcal{L}_{\rm DNB}$ & TS$_{\rm DM}^{\rm DNB}$ & TS$_{\rm B/N}^{\rm DNB}$ & $\Delta\log\mathcal{L}_{\rm NDM}$ & TS$_{\rm DM}^{\rm NDM}$ & TS$_{\rm N}^{\rm NDM}$ \\
\hline
Cholis21\_8l & -2148 & 12772 & -513 & 4143/3677 & 18 & 11749 & 15/124 & 15 & 11997 & 123 \\
Cholis21\_ch & -1664 & 14091 & -500 & 5400/4203 & 82 & 8850 & 632/328 & 8 & 13691 & 51 \\
Cholis21\_bf & -1562 & 15141 & -681 & 5223/4189 & 53 & 10871 & 373/239 & 8 & 14757 & 52 \\
DiMauro21 & -2649 & 17198 & -566 & 4024/10369 & 48 & 15259 & 2/566 & 48 & 15338 & 563 \\
Macias19 & -1718 & 13231 & -109 & 6248/5352 & 132 & 7179 & 1769/625 & 10 & 12485 & 101 \\
Pohl22 & -2195 & 6137 & 343 & 6030/5519 & 347 & 764 & 3952/3704 & 121 & 4432 & 2763 \\
Galp21\_SA0 & -918 & 19875 & -131 & 9271/8207 & 201 & 8535 & 1988/1392 & 31 & 16889 & 429 \\
Galp21\_SA50 & -937 & 22909 & -256 & 9634/8770 & 194 & 10104 & 1639/1349 & 37 & 21073 & 353 \\
Galp21\_SA100 & 0 & 19294 & -242 & 8135/9733 & 124 & 9928 & 1150/2194 & 30 & 17714 & 718 \\
\hline
\end{tabular}
\caption{Summary of the goodness of fit and test statistics for the DM, NB, and BB templates under different IEM choices. 
Column~(2) reports $\Delta\log\mathcal{L}_{\rm DM}\equiv \log\mathcal{L}_{\rm DM}-\log\mathcal{L}_{\rm DM}^{\max}$ with respect to the best-fitting IEM (\texttt{Galp21\_SA100}). 
Column~(3) gives TS$_{\rm DM}$ for the DM-only fit. 
Columns~(4) and (5) report the change in log-likelihood and the TS values when replacing DM with NB+BB (TS$_{\rm B/N}$ is reported as BB/NB). 
Columns~(6)--(8) report the improvement $\Delta\log\mathcal{L}_{\rm DNB}$ and the corresponding TS values when adding NB+BB on top of DM (TS$_{\rm B/N}^{\rm DNB}$ is reported as BB/NB). 
Columns~(9)--(11) report the change $\Delta\log\mathcal{L}_{\rm NDM}$ and the TS values for the fit including DM+NB only (TS$_{\rm N}^{\rm NDM}$ refers to the NB template). 
Negative $\Delta\log\mathcal{L}$ values indicate a worse fit relative to the corresponding reference model.}
\label{tab:E_summary}
\end{table*}


\emph{Spatial distribution.—}
To obtain a minimally model-dependent characterization of the GCE morphology, we extract its radial profile by fitting a set of concentric annuli whose normalizations are allowed to vary independently.
Denoting by $S_i$ the best-fit energy flux assigned to annulus $i$ and by $\Delta\Omega_i$ its solid angle, we define the corresponding surface brightness as $dS/d\Omega(\theta_i)\equiv S_i/\Delta\Omega_i$,
where $\theta_i$ is the characteristic angular distance of the annulus from the GC line of sight.
This procedure provides a flexible, template-agnostic reconstruction of the angular profile. We show here the results obtained when considering the \textbf{Benchmark} setup with $1-10$ GeV data but as shown in Appendix \ref{app:results} we obtain similar results for the other analysis setups.

\begin{figure}
  \centering
  \includegraphics[width=0.99\linewidth]{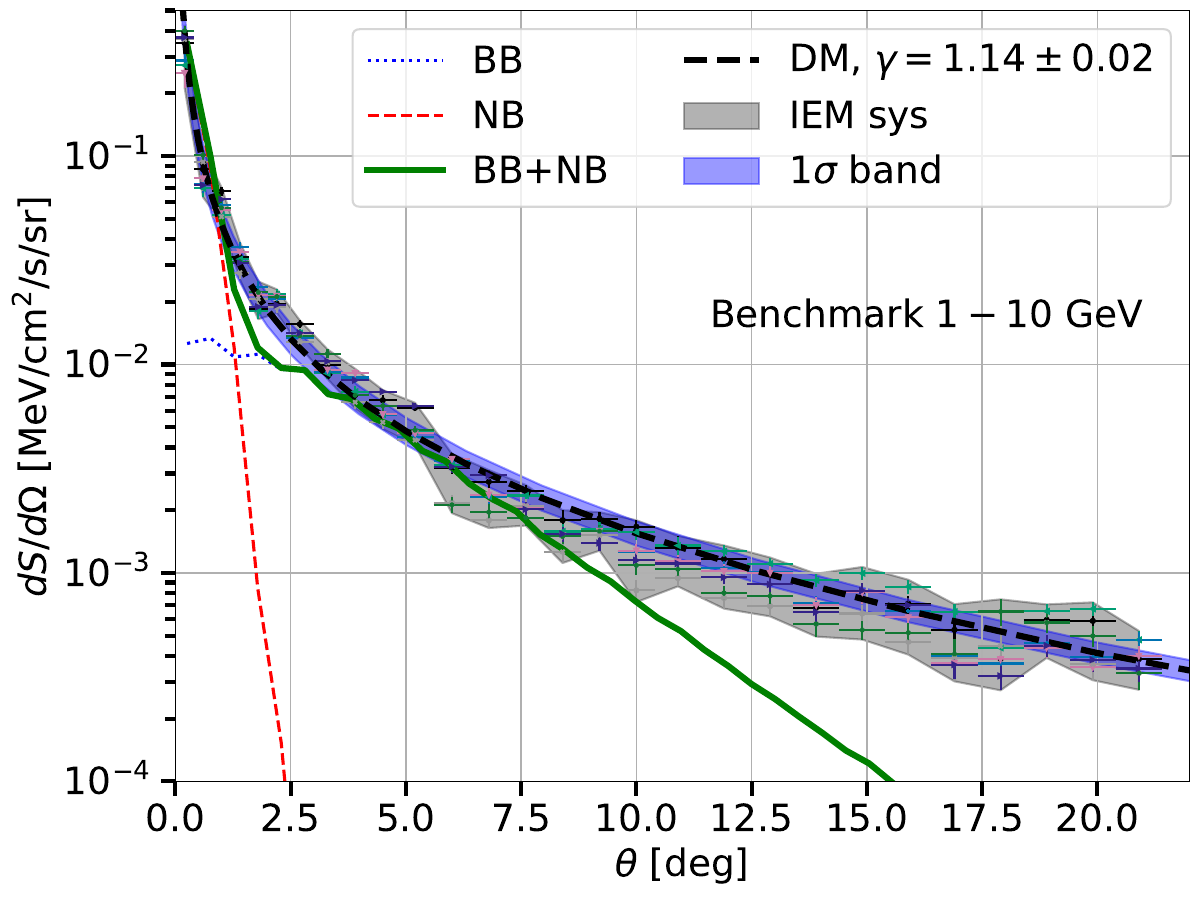}
  \caption{Surface brightness, $dS/d\Omega$, as a function of angular distance from the GC line of sight, reconstructed using the set of IEMs adopted in this work (colored points). The black dashed curve shows the best-fit gNFW profile ($\gamma=1.15$) with its $1\sigma$ uncertainty band (blue). The gray band indicates the spread associated with the IEM choice. The red dashed and blue dotted curves show the NB and BB components obtained from a fit using only NB+BB templates, with their sum shown in green.}
  \label{fig:SB}
\end{figure}

Fig.~\ref{fig:SB} shows the reconstructed $dS/d\Omega$ obtained with each of the IEMs considered in this work.
In agreement with previous analyses (see, e.g., Ref.~\cite{DiMauro:2021prd}), the surface brightness is well described over the full angular range by a generalized NFW (gNFW) template, with best-fit inner slope $\gamma = 1.15 \pm 0.02$.
The quoted uncertainty includes both the statistical component and the systematic spread induced by the IEM choice.

For comparison, Fig.~\ref{fig:SB} also reports the surface brightness implied by a fit in which the excess is modeled exclusively with the NB and BB templates of Ref.~\cite{Coleman:2019kax}.
As expected, NB dominates the innermost region ($\theta\lesssim 1^\circ$) but declines rapidly with angle.
Beyond $\theta\gtrsim 2^\circ$, BB becomes dominant; however, the transition region where the NB-to-BB handover occurs yields a summed profile that is about a factor of 2 smaller than the measured GCE surface brightness.
At larger angles, the data exhibit a comparatively shallow decline relative to the bulge components, such that the NB+BB prediction undershoots the reconstructed profile for $\theta \gtrsim 10^\circ$. In particular, at $\theta=15^{\circ}$ the NB+BB prediction is about a factor of 10 smaller than the measured GCE $dS/d\Omega$.
These discrepancies persist when accounting for the envelope associated with the IEM choice, indicating that a pure NB+BB interpretation does not reproduce the full radial morphology extracted from the data in our analysis.
If we try to fit the surface brightness data using the NB and the DM templates together, the fit prefers to reduce significantly the contribution of the former component because the DM contribution provides a much better fit if considered alone. Ref.~\cite{McDermott:2022zmq} found similar conclusions for large angular distances from the GC.

\begin{figure}
  \centering
  \includegraphics[width=0.99\linewidth]{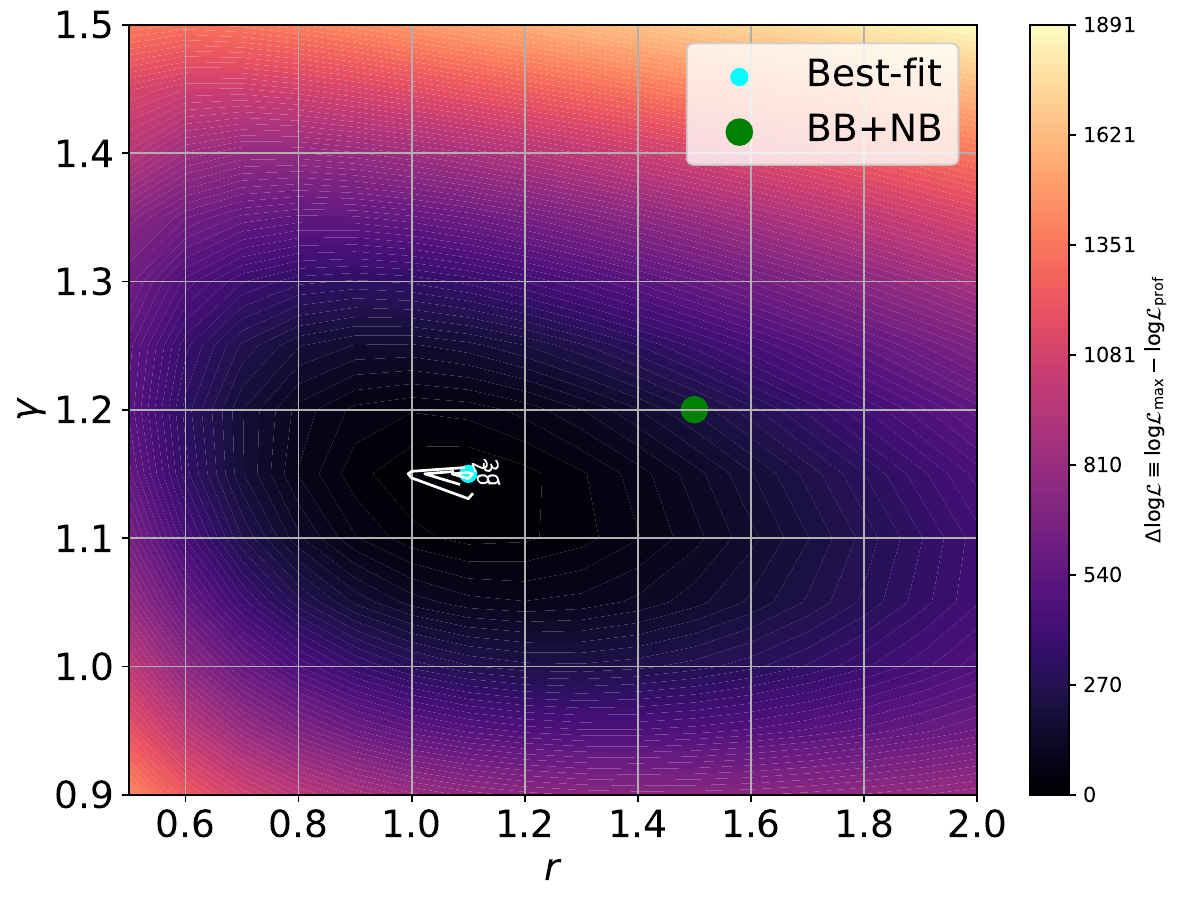}
  \caption{Profiled log-likelihood difference $\Delta\log\mathcal{L}$ as a function of the gNFW inner slope $\gamma$ and axis ratio $r$ (elongation along the Galactic plane relative to the perpendicular direction), obtained after profiling over the IEM choice. The marker indicates the best-fit point, the white contours show the $1/2/3\sigma$ regions, and the green marker indicates the location corresponding to the NB+BB morphology.}
  \label{fig:shape}
\end{figure}

We further test whether the GCE is consistent with spherical symmetry or exhibits an elongation along the Galactic plane.
Such a departure from sphericity is expected for the BB template~\cite{Coleman:2019kax}, and has also been discussed for the DM distribution in recent N-body simulations~\cite{Grand:2022olu,Muru:2025vpz}.
Following these studies, we parametrize departures from sphericity through an axis ratio $r$, defined as the ratio of the isointensity axis along the Galactic plane to that perpendicular to it.
Typical expectations are $r\simeq 1.1$--$1.5$ for DM-like halos and $r\simeq 1.5$ for NB+BB-like morphologies.
We fit the data with a gNFW template scanning both the inner slope $\gamma$ and the axis ratio $r$ over the range $r\in[0.5,2.0]$.
After profiling over the IEM choice, we find $\gamma=1.15\pm0.02$ and $r=1.10\pm0.05$.
Fig.~\ref{fig:shape} shows the profiled $\Delta\log\mathcal{L}$ as a function of $(\gamma,r)$.
The preferred morphology is therefore consistent with an approximately spherical profile, with at most a mild elongation along the Galactic plane, compatible with the trends suggested by DM in Refs.~\cite{Grand:2022olu,Muru:2025vpz}.

\begin{figure}
  \centering
  \includegraphics[width=0.99\linewidth]{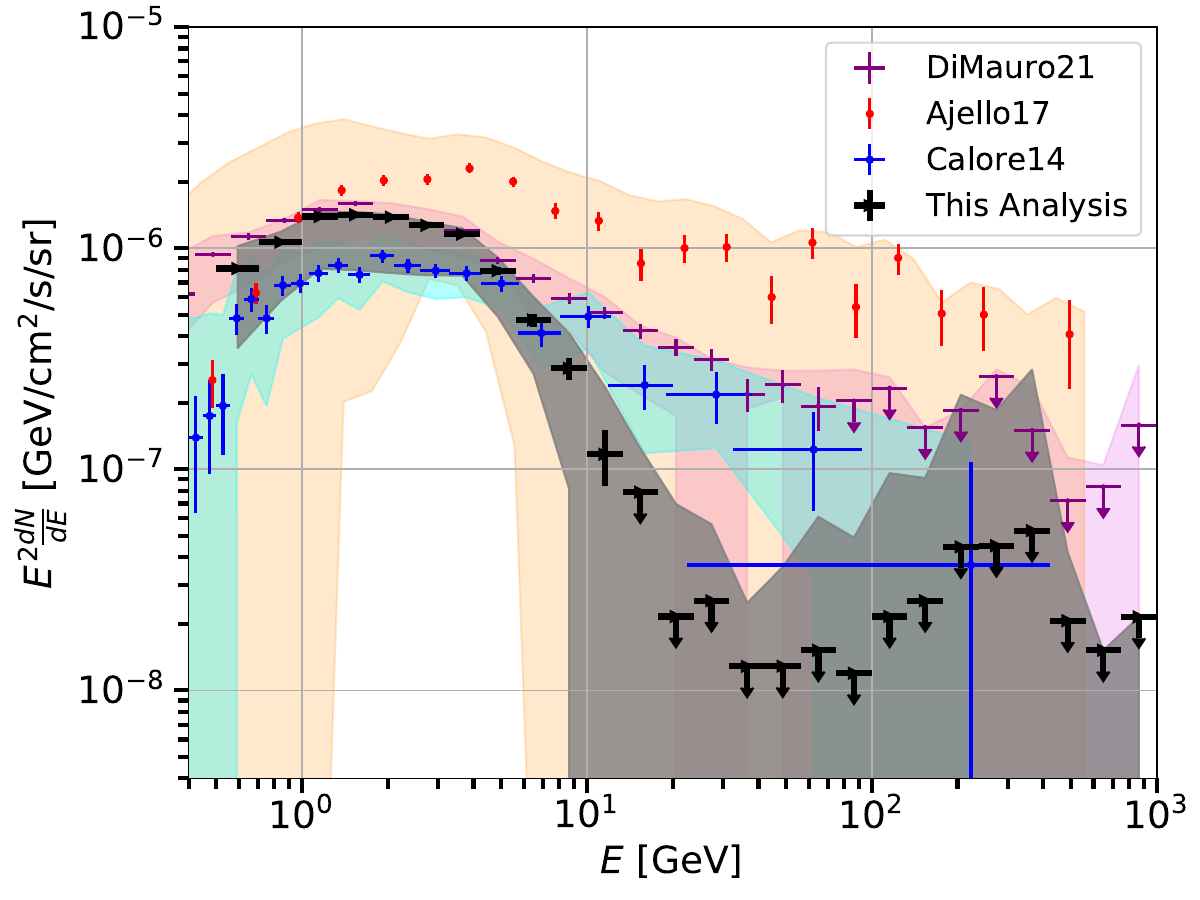}
  \caption{GCE SED obtained with our analysis compared with previous papers. In particular, we display our result obtained with the {\tt Galp21\_SA100} and the envelope of all the spectra obtained with the different IEMs. We also show the flux reported in Refs.~\cite{Calore:2014nla,Fermi-LAT:2017opo,DiMauro:2021prd}}
  \label{fig:SED}
\end{figure}

\emph{Energy spectrum.—}
Having established that the GCE morphology is well described by a spherical gNFW template with inner slope $\gamma\simeq 1.15$, we fix this spatial template and extract the differential energy spectrum through a bin-by-bin likelihood analysis over the range $0.5$--$1000$~GeV.
The resulting spectral energy distribution (SED) is shown in Fig.~\ref{fig:SED}, where we compare the SEDs obtained with the different IEMs to previous determinations in Refs.~\cite{Calore:2014nla,Fermi-LAT:2017opo,DiMauro:2021prd}.
Once the IEM-driven spread is taken into account, our results are consistent with earlier measurements but with a much narrower systematic band.
The spectrum peaks at a few GeV and exhibits both low- and high-energy tails.
Quantitatively, the flux at $10$~GeV ($100$~GeV) is suppressed by a factor $\sim 10$ ($\sim 100$) relative to the peak value.

A key feature of the reconstructed SED is the rapid attenuation above $\sim 10$~GeV: the GCE emission becomes negligible and is constrained by upper limits at the level of $\sim 10^{-8}$--$10^{-7}$~GeV\,cm$^{-2}$\,s$^{-1}$\,sr$^{-1}$.
This behavior is potentially informative for astrophysical interpretations.
In particular, if the GCE originates from a bulge population of MSPs, inverse-Compton emission from the associated $e^\pm$ population scattering on the ISRF is expected to contribute at $\gtrsim 10$~GeV.
The absence of a comparable high-energy component in Fig.~\ref{fig:SED} may therefore require a reduced $e^\pm$ yield (or a different injection spectrum and/or transport environment) for bulge MSPs relative to their disk counterparts; we will quantify this requirement in a dedicated follow-up study (see also Ref.~\cite{2021MNRAS.507.5161S}).

Conversely, as discussed in a number of previous works~\cite{Hooper:2010mq,Calore:2014xka,DiMauro:2021qcf,Koechler:2025ryv,Kong:2025ccv}, the measured GCE spectrum is compatible with annihilating DM in hadronic, leptonic, or mixed final states for $m_{\rm DM}\sim 30$--$60$~GeV and an annihilation cross section close to the thermal benchmark.

\emph{Conclusions.—}
In this Letter we have presented a new analysis of the GCE that achieves a substantially improved agreement between \emph{Fermi}-LAT data and models based on different IEMs and analysis techniques.
Relative to previous studies, which typically exhibited fractional residuals at the $30$--$40\%$ level~\cite{Cholis:2021rpp,DiMauro:2021prd,Pohl:2022nnd}, our optimized pipeline reduces the residuals to $\lesssim 10\%$ over most of the ROI.
This improvement enables tighter control of the dominant systematic uncertainties associated with diffuse-emission modeling and therefore a more precise determination of the GCE properties.
We find that the GCE morphology is well described by an approximately spherical gNFW template, with inner slope $\gamma \sim 1.15$ and axis ratio $r \sim1.10$.
We further show that modeling the excess exclusively with the bulge-tracing templates (BB+NB) fails to reproduce the reconstructed surface-brightness profile at line-of-sight angles of about $2^\circ$ and $>10^\circ$.
Moreover, irrespective of the adopted IEM and even when BB and NB are included simultaneously in the fit, the DM-motivated component remains highly significant, with a nominal significance in the range $\sim 28$--$120\sigma$.
These conclusions are robust to a number of cross-checks aimed at reducing sensitivity to Galactic-plane mismodeling, including masking the Galactic plane, restricting the energy range to $1$--$10$~GeV, and applying the weighted-likelihood technique.

Finally, we provide an updated measurement of the GCE energy spectrum, which peaks at a few GeV and becomes negligible above $\sim 10$~GeV, with upper limits at the level of $E^2\Phi \lesssim \rm{few} \,10^{-8}$~GeV\,cm$^{-2}$\,s$^{-1}$\,sr$^{-1}$ for all the IEM models tested.
This spectra shape, that is perfectly compatible with DM particles with mass around $30-60$ GeV with a thermal cross section, instead strongly constrains a possible high-energy IC component from a bulge MSP population; a quantitative interpretation in terms of MSP $e^\pm$ injection will be presented in a forthcoming study.
We will also present a dedicated assessment of the DM hypothesis using state-of-the-art simulation inputs as in Ref.~\cite{Muru:2025vpz}. Stay tuned...

\begin{acknowledgments}
The {\it Fermi} LAT Collaboration acknowledges generous ongoing support from a number of agencies and institutes that have supported both the development and the operation of the LAT as well as scientific data analysis. These include the National Aeronautics and Space Administration and the Department of Energy in the United States, the Commissariat\'a l'Energie Atomique and the Centre National de la Recherche Scientifique / Institut National de Physique Nucl\'eaire et de Physique des Particules in France, the Agenzia Spaziale Italiana and the Istituto Nazionale di Fisica Nucleare in Italy, the Ministry of Education, Culture, Sports, Science and Technology (MEXT), High Energy Accelerator Research Organization (KEK) and Japan Aerospace Exploration Agency (JAXA) in Japan, and the K. A. Wallenberg Foundation, the Swedish Research Council and the Swedish National Space Board in Sweden.
Additional support for science analysis during the operations phase is gratefully acknowledged from the Istituto Nazionale di Astrofisica in Italy and the Centre National d'Etudes Spatiales in France. This work performed in part under DOE Contract DE- AC02-76SF00515.

M.D.M. thanks the Fermi-LAT team for providing comments during the internal review process that improved the paper.
M.D.M. acknowledges support from the research grant {\sl TAsP (Theoretical Astroparticle Physics)} funded by Istituto Nazionale di Fisica Nucleare (INFN) and from the Italian Ministry of University and Research (MUR), PRIN 2022 ``EXSKALIBUR – Euclid-Cross-SKA: Likelihood Inference Building for Universe’s Research'', Grant No. 20222BBYB9, CUP I53D23000610 0006, and from the European Union -- Next Generation EU.
\end{acknowledgments}

\newpage

\appendix

\onecolumngrid

\newpage

\section{Interstellar-emission models}
\label{app:IEMs}

Accurate modeling of the Galactic interstellar emission (IEM) is essential for any analysis of the inner Galaxy, where diffuse foregrounds dominate the observed $\gamma$-ray intensity and drive the leading systematic uncertainties \cite{Calore:2014xka,TheFermi-LAT:2017vmf,Leane:2019uhc,Chang:2019ars,Zhong:2019ycb,Calore:2021jvg,List:2025qbx}.
To assess the robustness of our results against diffuse-modeling assumptions, we repeat the analysis using a suite of IEMs that have been developed and/or optimized for Galactic-center studies.
We refer to the original references~\cite{DiMauro:2021prd,Cholis:2021rpp,Porter:2021tlr,Coleman:2019kax,Pohl:2022nnd} for full technical details and summarize here the common framework and the principal differences among the adopted models.

\subsection{Common framework and components}
All IEMs used in this work are based on the \texttt{GALPROP} formalism~\cite{Porter:2021tlr}, which predicts diffuse $\gamma$-ray emission by numerically solving the CR transport equation for specified inputs (CR source distribution and injection spectra, propagation setup, and interstellar target fields).
In all cases, the diffuse emission is decomposed into the standard physically motivated components:
(i) inverse-Compton (IC) emission from CR electrons scattering on the interstellar radiation field (ISRF), typically provided as separate contributions from the CMB, starlight, and infrared target fields;
(ii) bremsstrahlung from CR leptons interacting with interstellar gas; and
(iii) $\pi^0$-decay emission from hadronic CR interactions with interstellar medium atoms.
In addition, all model families include large-scale templates that can impact the inner Galaxy, most notably the \emph{Fermi} bubbles and prominent local structures such as Loop~I, as well as the isotropic background and the solar and lunar emission.
In the maximum likelihood maximization we leave free to vary for each IEM component the normalization and a change of the spectral shape.
This applies also to the the isotropic background for which we allow a spectral rescaling rather than adopting a fixed spectral form as it is typically done in standard catalog analyses.\footnote{https://fermi.gsfc.nasa.gov/ssc/data/access/lat/BackgroundModels.html}

Several IEM families further employ a Galactocentric ring (annulus) decomposition, in which gas- and/or IC-correlated emission is provided in multiple rings in Galactocentric radius.
This approach allows the fit to adjust the relative normalizations of different line-of-sight regions, partially absorbing uncertainties from gas maps, emissivities, source distributions while preserving a physically interpretable component structure.
Other models adopt a more compact description where each physical channel is represented by a small number of all-sky templates (e.g., one bremsstrahlung, one $\pi^0$-decay, and three IC templates).

Across this common backbone, the models differ primarily in: (a) the assumed spatial distribution of CR sources; (b) the construction, dust correction, and kinematic decomposition of gas maps (H\,\textsc{i} and H$_2$/CO); (c) the adopted ISRF model and hence the IC morphology; (d) the propagation setup and (in some cases) tuning to local CR data; and (e) the treatment of large-scale structures (bubbles, Loop~I) and additional astrophysically motivated templates used in GC studies.

\subsection{IEM sets adopted in this work}
\begin{itemize}
  \item \textbf{DiMauro21}~\cite{DiMauro:2021prd}.
  We adopt the baseline IEM used in Ref.~\cite{DiMauro:2021prd}, constructed within the \texttt{GALPROP} framework and based on the Yusifov pulsar source distribution~\cite{Yusifov:2004fr}.
  In our implementation this is a compact template set, consisting of one bremsstrahlung and one $\pi^0$-decay map, three IC templates (CMB, starlight, IR), and two bubble templates (inner/outer), for a total of 7 diffuse templates.

  \item \textbf{Cholis21}~\cite{Cholis:2021rpp}.
  We include the three best-performing models from Ref.~\cite{Cholis:2021rpp} (labels \texttt{8l}, \texttt{ch}, \texttt{bf}), selected for their ability to reproduce local primary and secondary CR data.\footnote{We also tested the \texttt{8t} and \texttt{c4} models, which are among the top five in Ref.~\cite{Cholis:2021rpp}, but they yield a substantially worse fit than the other realizations and are therefore not used.}
  This ensemble provides a controlled bracket of propagation/source assumptions constrained by CR measurements, while retaining a compact diffuse-template structure analogous to \textbf{DiMauro21} (same set of physical components and template counting in our implementation), with the exception that the IC emission is provided as a single template.

  \item \textbf{Galp21}~\cite{Porter:2021tlr}.
  We consider three representative IEMs distributed with the public \texttt{GALPROP} release of Ref.~\cite{Porter:2021tlr}, which adopt an ISRF model based on Ref.~\cite{2012A&A...545A..39R}.
  These models provide the diffuse channels in a ring-decomposed form; in our implementation the splitting of IC and gas-correlated emission into multiple annuli yields $\mathcal{O}(20)$ diffuse templates.
  We use the three configurations labeled \texttt{SA0}, \texttt{SA50}, and \texttt{SA100}, which probe different CR source prescriptions: \texttt{SA0} corresponds to an axisymmetric baseline (including commonly used SNR/pulsar-inspired distributions~\cite{1998ApJ...504..761C,Yusifov:2004fr}), whereas \texttt{SA50} and \texttt{SA100} incorporate a spiral-arm source component following Ref.~\cite{Porter:2017vaa}, with a 50/50 split between disk-like and spiral-arm injection for \texttt{SA50}, and a pure spiral-arm configuration for \texttt{SA100}.

  \item \textbf{Macias19}~\cite{Coleman:2019kax}.
  We adopt the IEM of Ref.~\cite{Coleman:2019kax}, which emphasizes an improved dust-based correction in the construction of the gas templates and includes an explicit treatment of positive/negative dust residual contributions.
  This model is ring-decomposed (in our implementation: 6 IC rings and 4 gas rings, for a total of 19 diffuse templates) and is often used in conjunction with stellar-bulge templates (nuclear and boxy/peanut bulge) to trace a putative MSP population in the inner Galaxy.

  \item \textbf{Pohl22}~\cite{Pohl:2022nnd}.
  We employ the updated IEM presented in Ref.~\cite{Pohl:2022nnd}, which builds on the \textbf{Macias19} framework but implements a revised reconstruction of the inner-Galaxy H\,\textsc{i} distribution.
  This update directly impacts the morphology of gas-correlated $\pi^0$-decay and bremsstrahlung emission in the central few kiloparsecs, providing a targeted probe of one of the dominant systematics in GC analyses (line-of-sight gas assignment and ring decomposition).
  The overall ring/template structure is otherwise the same as in \textbf{Macias19} in our implementation.
\end{itemize}

We adopted the IEM set described above for two main reasons. First, all models are constructed within the \texttt{GALPROP} framework, which provides a physically grounded and self-consistent prediction of the diffuse $\gamma$-ray emission---including $\pi^0$-decay, bremsstrahlung, and IC components---by solving the CR transport equation given explicit assumptions on CR sources, propagation, interstellar gas, and the interstellar radiation field. Second, these IEM families have been widely used in recent GCE studies that reached different conclusions about the origin of the excess: in particular, the \textbf{Macias19} and \textbf{Pohl22} models have been employed in analyses favoring bulge-tracing templates and an MSP interpretation, whereas \textbf{DiMauro21} and \textbf{Cholis21} have been used in studies finding a more spherical, DM-compatible excess. Our aim is therefore to bracket the dominant diffuse-model systematics on the inferred GCE morphology and spectrum by adopting representative IEMs spanning the range of assumptions and conclusions present in the literature.

For the diffuse components, we allow independent normalizations and a spectral tilt parameter to vary in the likelihood fit.
Taken together, these IEM families span the principal, physically motivated axes of diffuse-model uncertainty relevant for the GCE---CR source gradients, propagation/tuning choices, ISRF-driven IC morphology, gas-map construction and ring assignment, and the handling of large-scale extended structures.
The specific realizations adopted in each family are summarized in Table~\ref{tab:IEM_summary}.

In this work we use the NB/BB maps as \emph{$\gamma$-ray intensity templates}, i.e.~as a proxy for prompt bulge-tracing emission (as expected, for instance, from unresolved MSP magnetospheric $\gamma$ rays).
However, if MSPs follow the bulge stellar distribution, they can also inject CR leptons (and possibly hadrons), whose propagated emission (IC for $e^\pm$, and $\pi^0$-decay/bremsstrahlung for nuclei/leptons) would generally have a morphology that differs from the underlying NB/BB stellar maps.
In particular, IC emission from bulge-injected $e^\pm$ is expected to be spatially broader because of diffusion and radiative cooling, and could in principle partially smooth the transition between NB- and BB-dominated regions.
Conversely, a bulge-shaped hadronic injection would mainly produce $\gamma$ rays correlated with the gas distribution and is therefore expected to be largely degenerate with the gas-related IEM components whose normalizations are floated in the fit.
A dedicated, fully physical MSP interpretation would thus require modeling injection and transport self-consistently (e.g.\ with \texttt{GALPROP}) rather than treating NB/BB purely as prompt templates.
Importantly, our measured SED and the stringent upper limits above $\sim 10$~GeV strongly constrain the allowed size of any additional IC component, suggesting that such a contribution—if present—must remain sub-dominant in our ROI; we defer a quantitative study to future work.

\begin{table*}[t]
\centering
\small
\setlength{\tabcolsep}{6pt}
\renewcommand{\arraystretch}{1.12}
\begin{tabular}{l l l}
\hline\hline
\textbf{IEM set} & \textbf{Assumption tested} & \textbf{Implementation in this work} \\
\hline
\multicolumn{3}{l}{\emph{Shared framework (all IEMs)}} \\
All &
\parbox[t]{0.26\textwidth}{Diffuse physical channels} &
\parbox[t]{0.62\textwidth}{\texttt{GALPROP}-based diffuse emission: IC (CMB/starlight/IR), bremsstrahlung, and $\pi^0$-decay.} \\
All &
\parbox[t]{0.26\textwidth}{Large-scale structures and standard backgrounds} &
\parbox[t]{0.62\textwidth}{Includes \emph{Fermi} bubbles and Loop~I as provided by each reference; plus mapcubes for isotropic emission, Loop~I, Sun, and Moon.} \\
All &
\parbox[t]{0.26\textwidth}{Fit degrees of freedom} &
\parbox[t]{0.62\textwidth}{For each diffuse component we float the normalization and an additional spectral tilt/index parameter.} \\
\hline
\multicolumn{3}{l}{\emph{Compact (channel-level) template sets}} \\
DiMauro21 &
\parbox[t]{0.26\textwidth}{CR source distribution; minimal template factorization} &
\parbox[t]{0.62\textwidth}{Baseline of Ref.~\cite{DiMauro:2021prd} (Yusifov pulsar source distribution). 7 diffuse templates: 1 bremsstrahlung, 1 $\pi^0$, 3 IC (CMB/SL/IR), and 2 bubbles (inner/outer); no ring decomposition.} \\
Cholis21 &
\parbox[t]{0.26\textwidth}{Propagation/source setups constrained by local CR data} &
\parbox[t]{0.62\textwidth}{Three best-performing realizations of Ref.~\cite{Cholis:2021rpp} (\texttt{8l, ch, bf}); implemented with the same compact 7-template structure as \textbf{DiMauro21}.} \\
\hline
\multicolumn{3}{l}{\emph{Ring-decomposed IEMs}} \\
Galp21\_SA0 &
\parbox[t]{0.26\textwidth}{Axisymmetric vs spiral-arm CR source prescription; IC morphology via ISRF} &
\parbox[t]{0.62\textwidth}{Public \texttt{GALPROP} release models of Ref.~\cite{Porter:2021tlr}. Ring-decomposed IC and gas-correlated emission, yielding $\mathcal{O}(20)$ diffuse templates in our implementation.} \\
Galp21\_SA50 &
\parbox[t]{0.26\textwidth}{Spiral-arm injection fraction} &
\parbox[t]{0.62\textwidth}{As \textbf{SA0}, but with spiral-arm sources and a 50/50 split of injected CR luminosity between disk-like and spiral-arm components.} \\
Galp21\_SA100 &
\parbox[t]{0.26\textwidth}{Pure spiral-arm injection} &
\parbox[t]{0.62\textwidth}{As \textbf{SA50}, but with injection entirely in spiral arms (pure spiral-arm configuration).} \\
Macias19 &
\parbox[t]{0.26\textwidth}{Gas maps and dust correction; dust residual treatment; ring freedom} &
\parbox[t]{0.62\textwidth}{GC-optimized IEM of Ref.~\cite{Coleman:2019kax} with improved dust-based gas corrections and explicit positive/negative dust residual templates. Ring decomposition with 6 IC rings and 4 gas rings; 19 diffuse templates total in our implementation.} \\
Pohl22 &
\parbox[t]{0.26\textwidth}{Inner-Galaxy H\,\textsc{i} reconstruction} &
\parbox[t]{0.62\textwidth}{Update of the Macias19 framework (Ref.~\cite{Pohl:2022nnd}) with a revised inner-Galaxy H\,\textsc{i} reconstruction; same ring/template structure and template counting as \textbf{Macias19}.} \\
\hline\hline
\end{tabular}
\caption{Summary of the decomposition, properties and characteristics of the IEMs used in this work and the dominant modeling axes they probe.}
\label{tab:IEM_summary}
\end{table*}

\section{Details of the analysis technique}
\label{app:analysis}

\begin{figure}
  \includegraphics[width=0.49\linewidth]{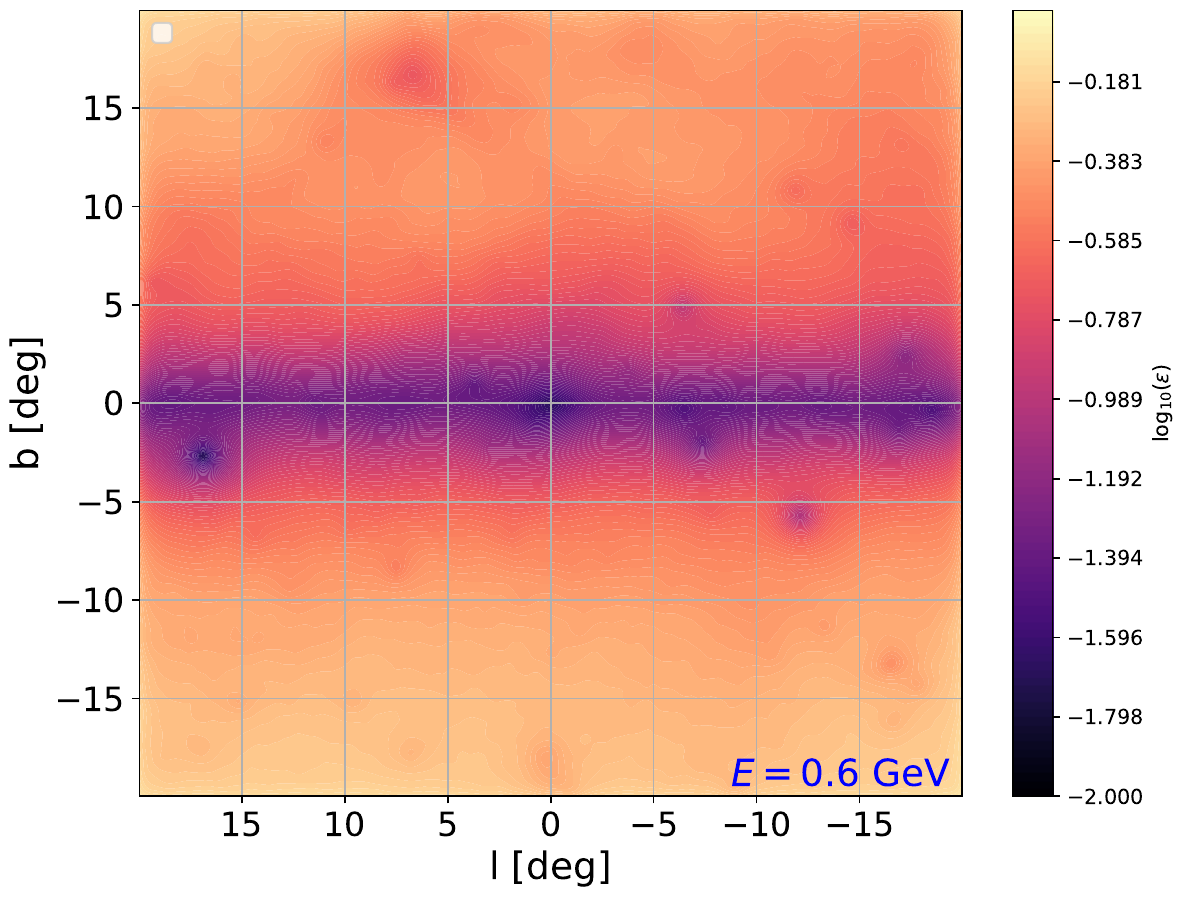}
  \includegraphics[width=0.49\linewidth]{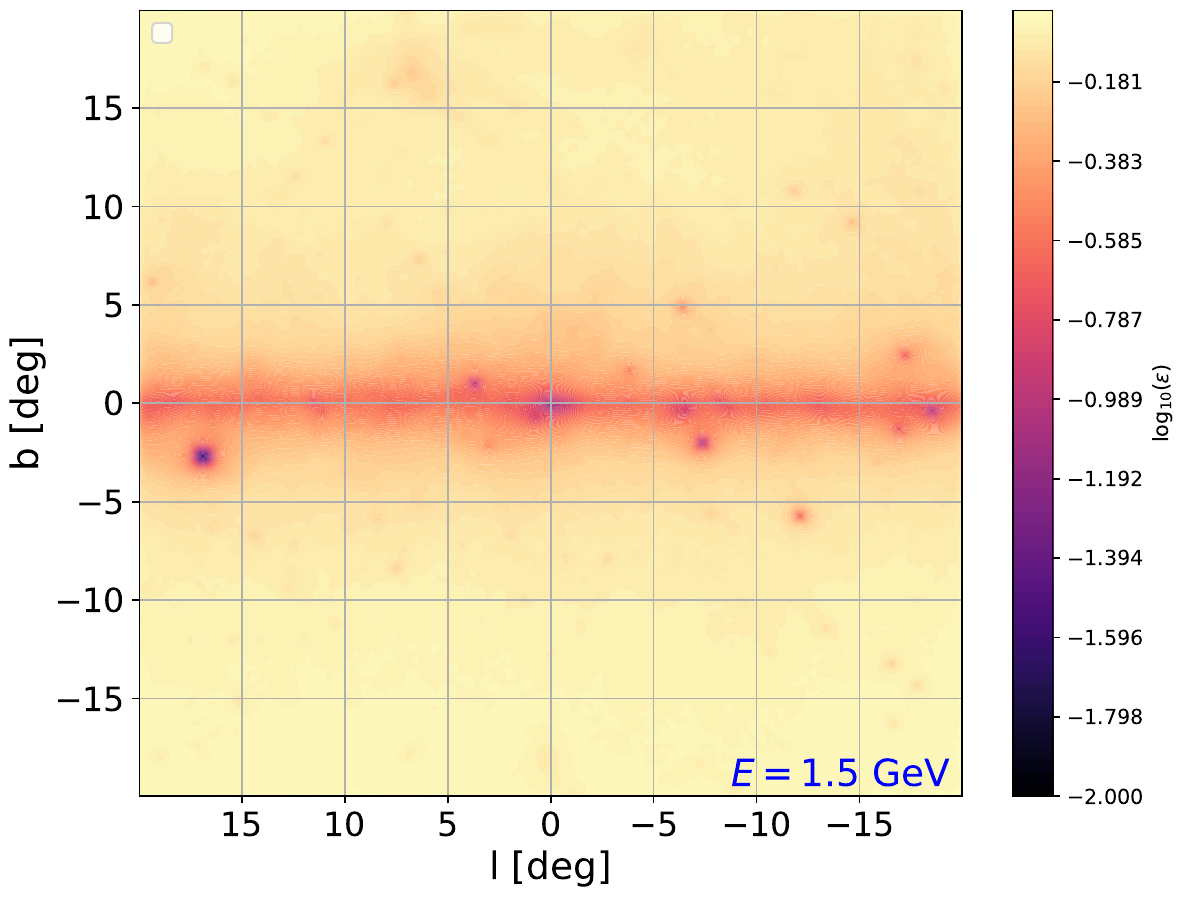}
  \caption{Weight maps for the weighted-likelihood analysis at $E=0.6$~GeV (left) and $E=1.5$~GeV (right) in the $40^\circ\times40^\circ$ GC ROI. The color scale shows $\log_{10}\epsilon$, where $\epsilon\in[0,1]$ is the per-bin weight.}
  \label{fig:weightmap}
\end{figure}

\begin{figure}
  \includegraphics[width=0.49\linewidth]{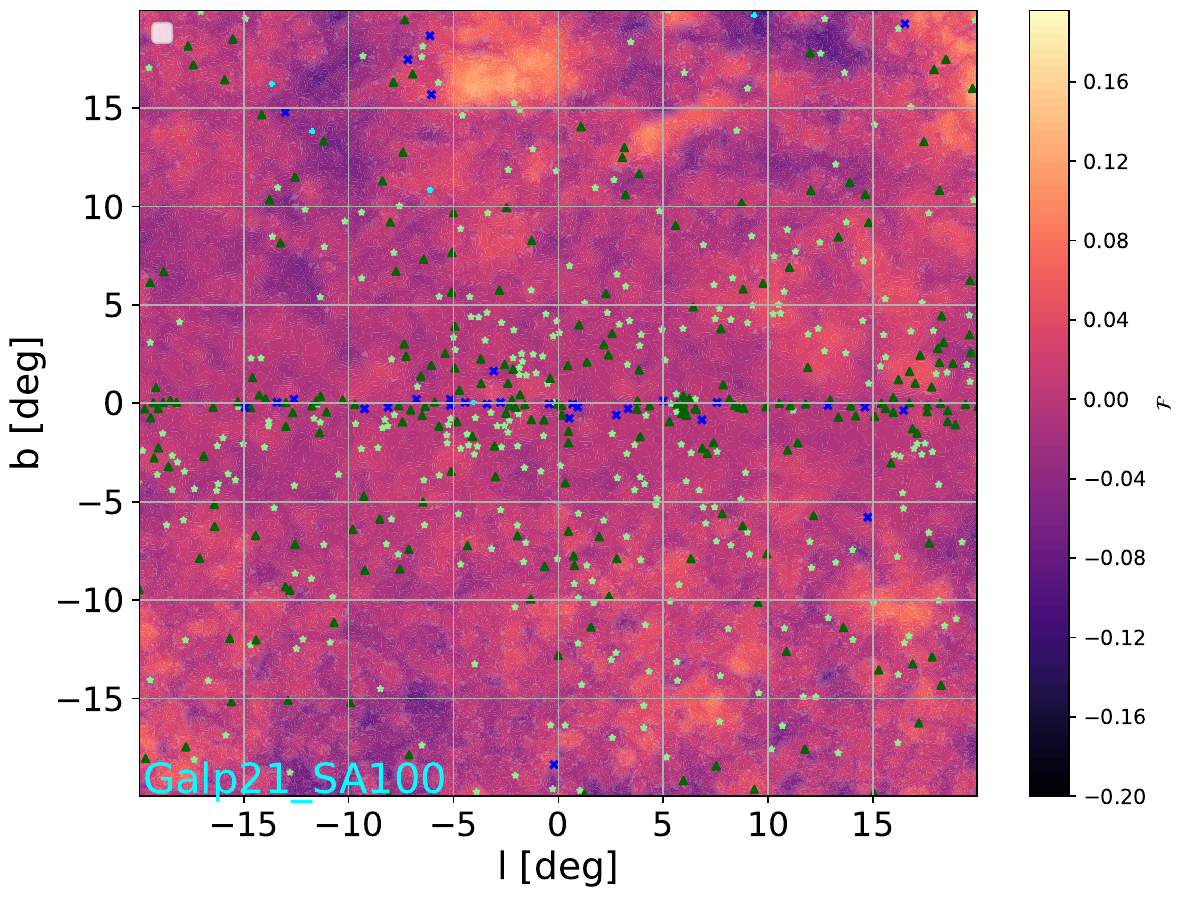}
  \includegraphics[width=0.49\linewidth]{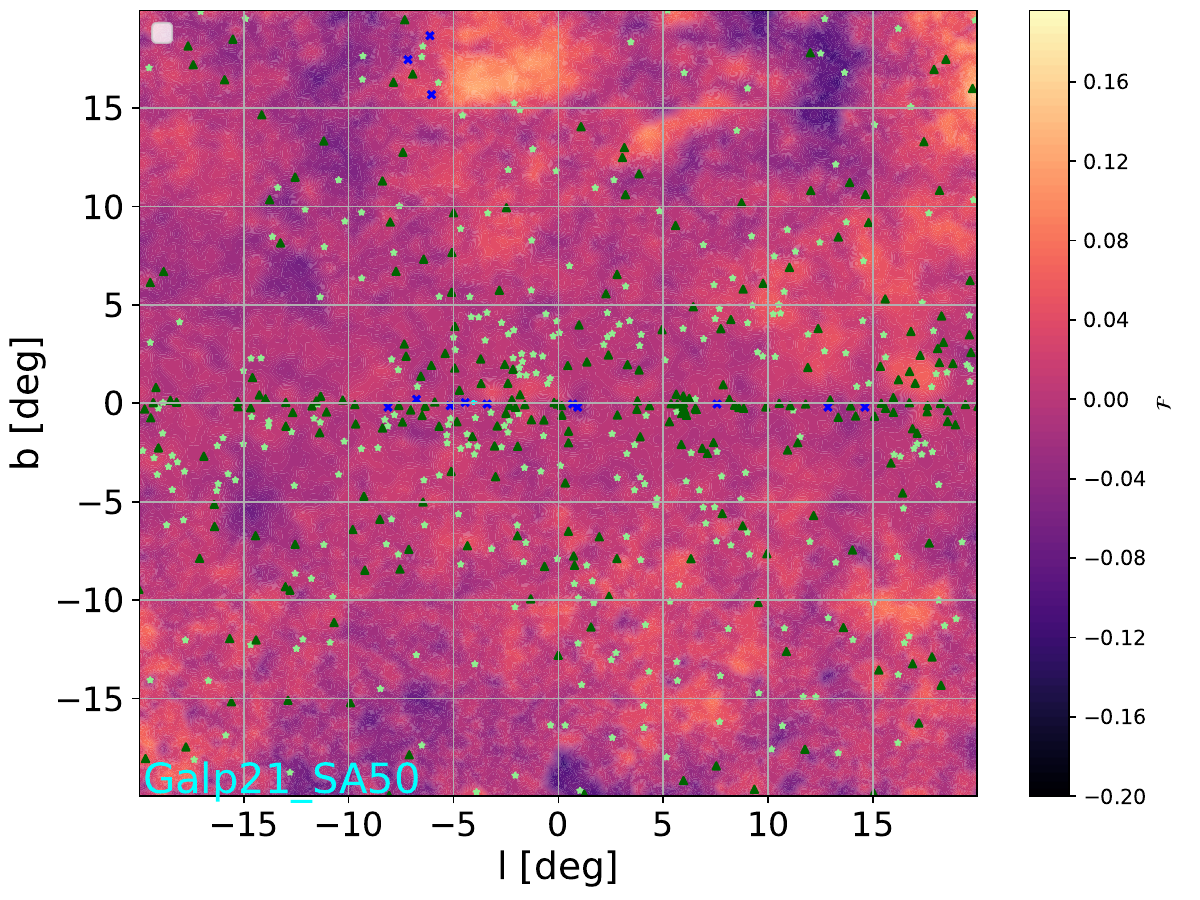}
  \includegraphics[width=0.49\linewidth]{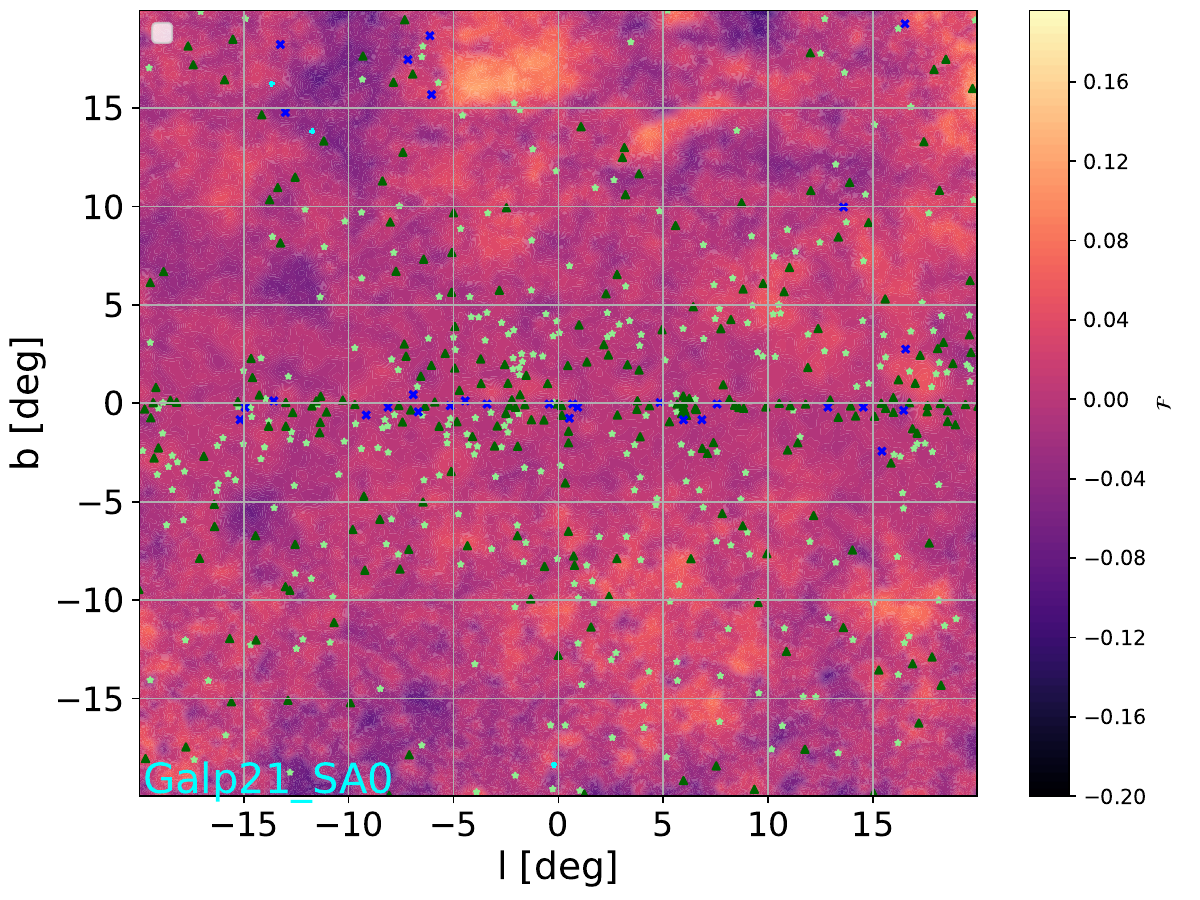}
  \includegraphics[width=0.49\linewidth]{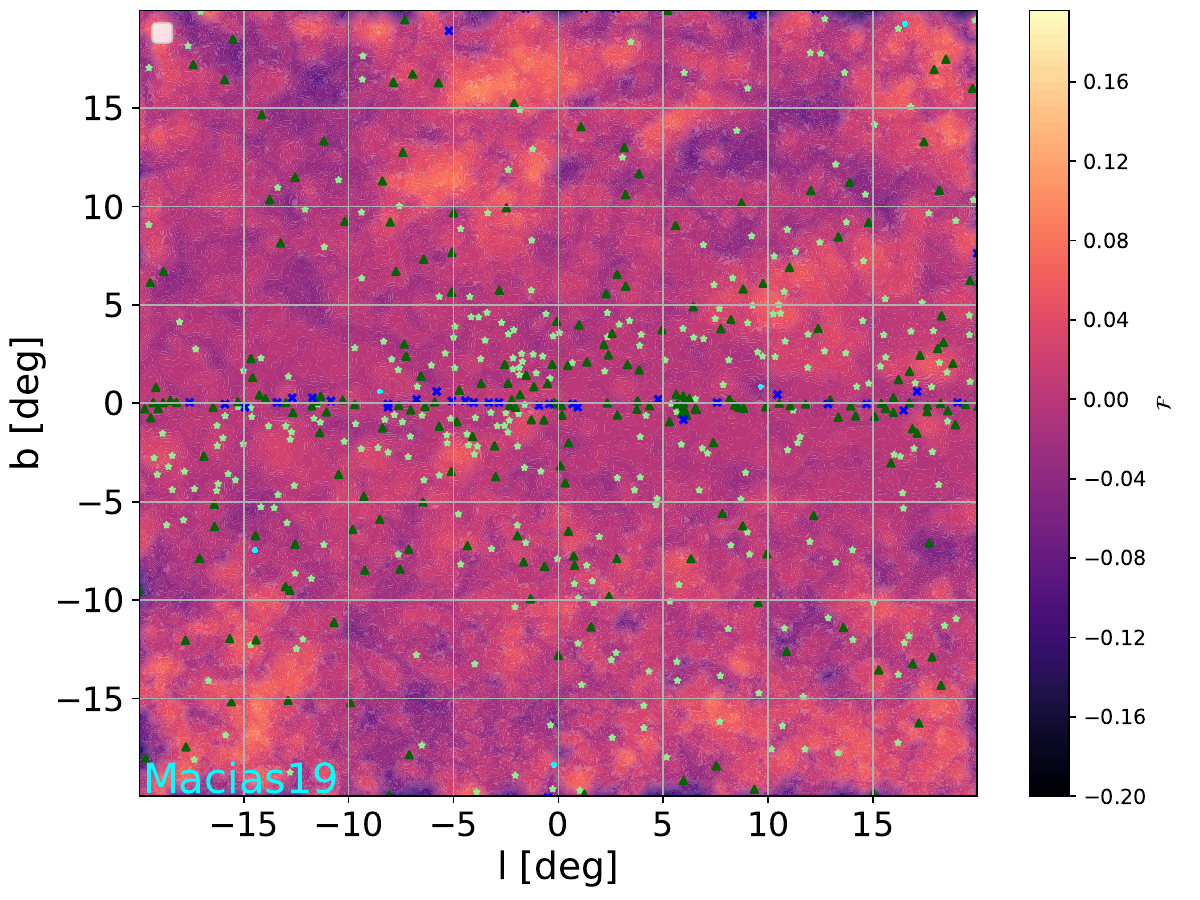}
  \includegraphics[width=0.49\linewidth]{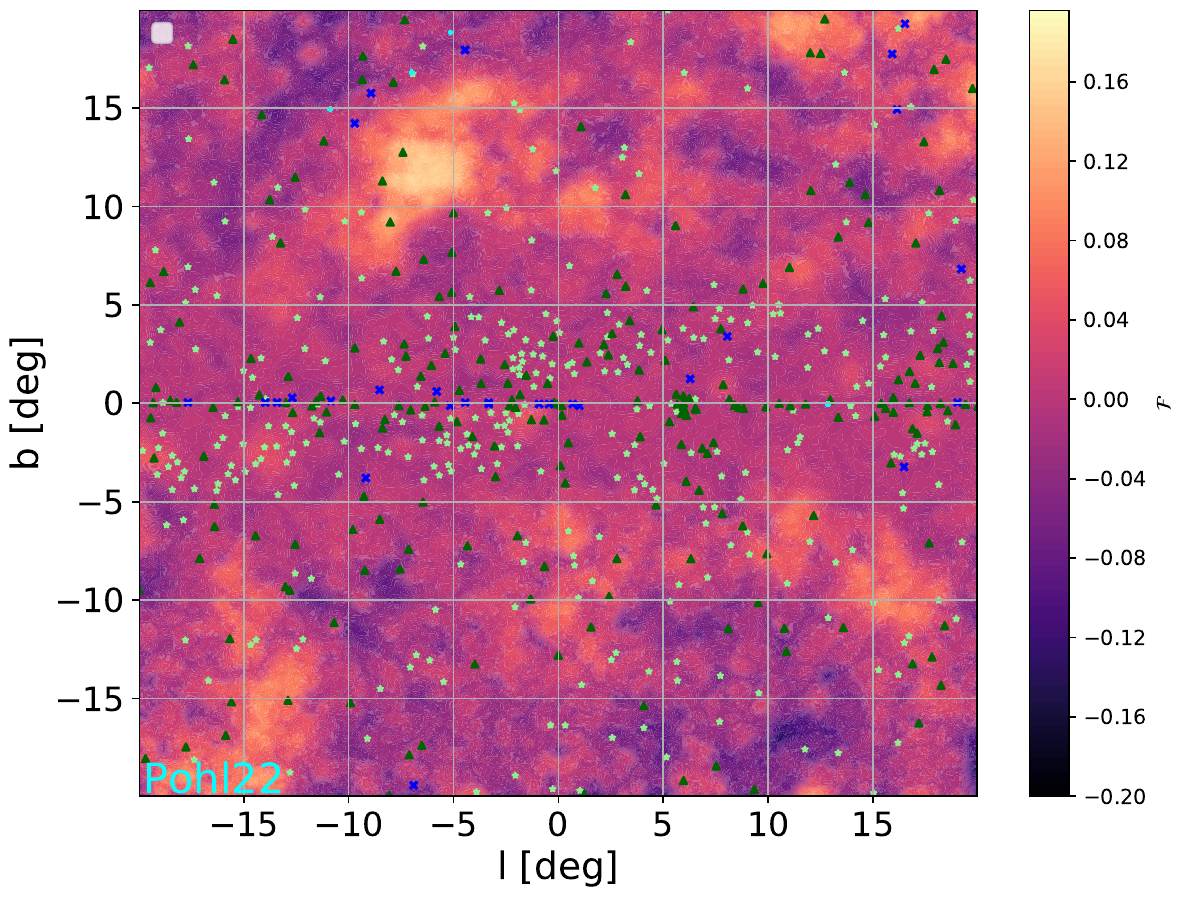}
  \includegraphics[width=0.49\linewidth]{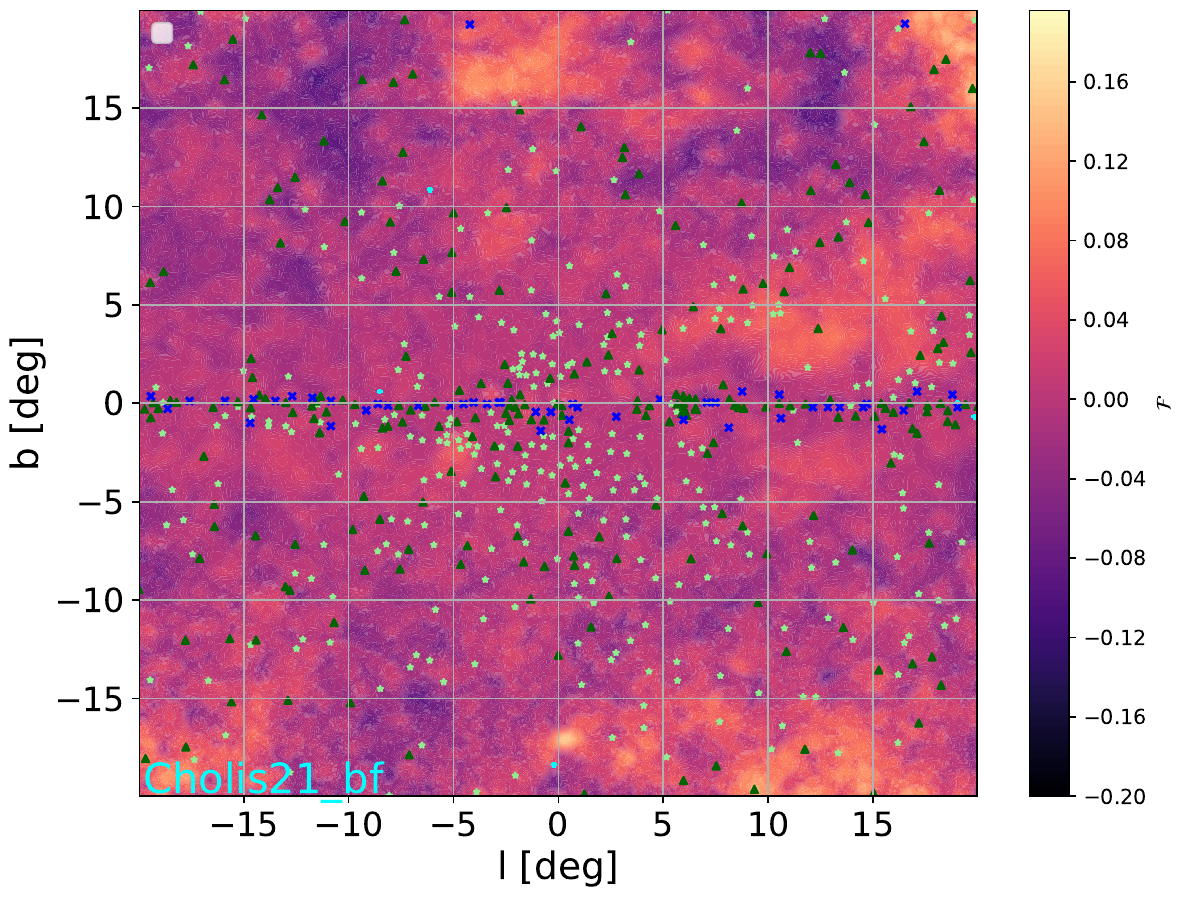}
  \caption{Fractional-residual maps in the $40^\circ\times 40^\circ$ ROI after the final refinement step for a representative subset of IEMs. Residuals are computed using $\mathcal{F}$ defined in Eq.~\ref{eq:Fapp}. We display the 4FGL sources detected with $TS<100$ (light green) and $TS>100$ (dark green) and newly detected sources (blue).}
  \label{fig:resmap}
\end{figure}

\begin{figure}
  \centering
  \includegraphics[width=0.49\linewidth]{fracmap_histo_40x40_cleaned.pdf}
  \includegraphics[width=0.49\linewidth]{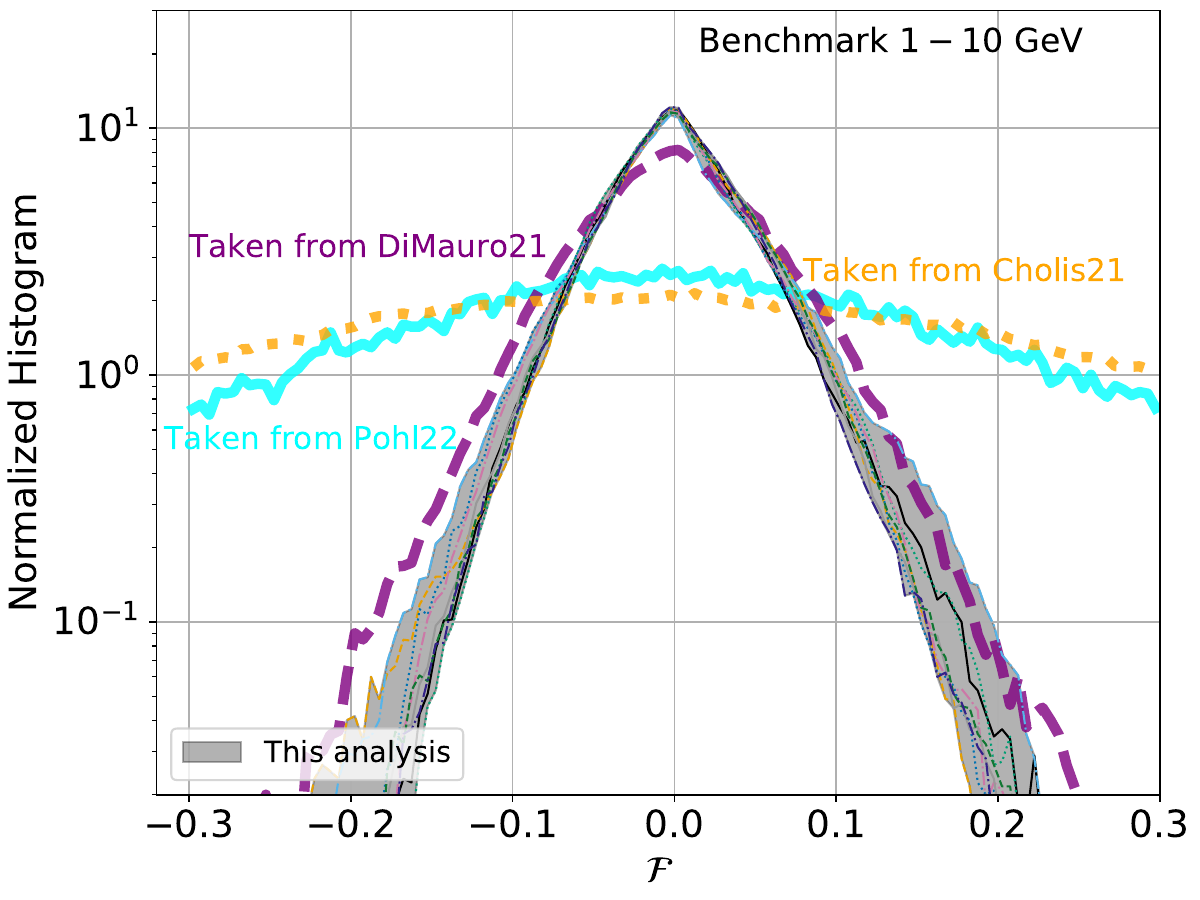}
  \includegraphics[width=0.49\linewidth]{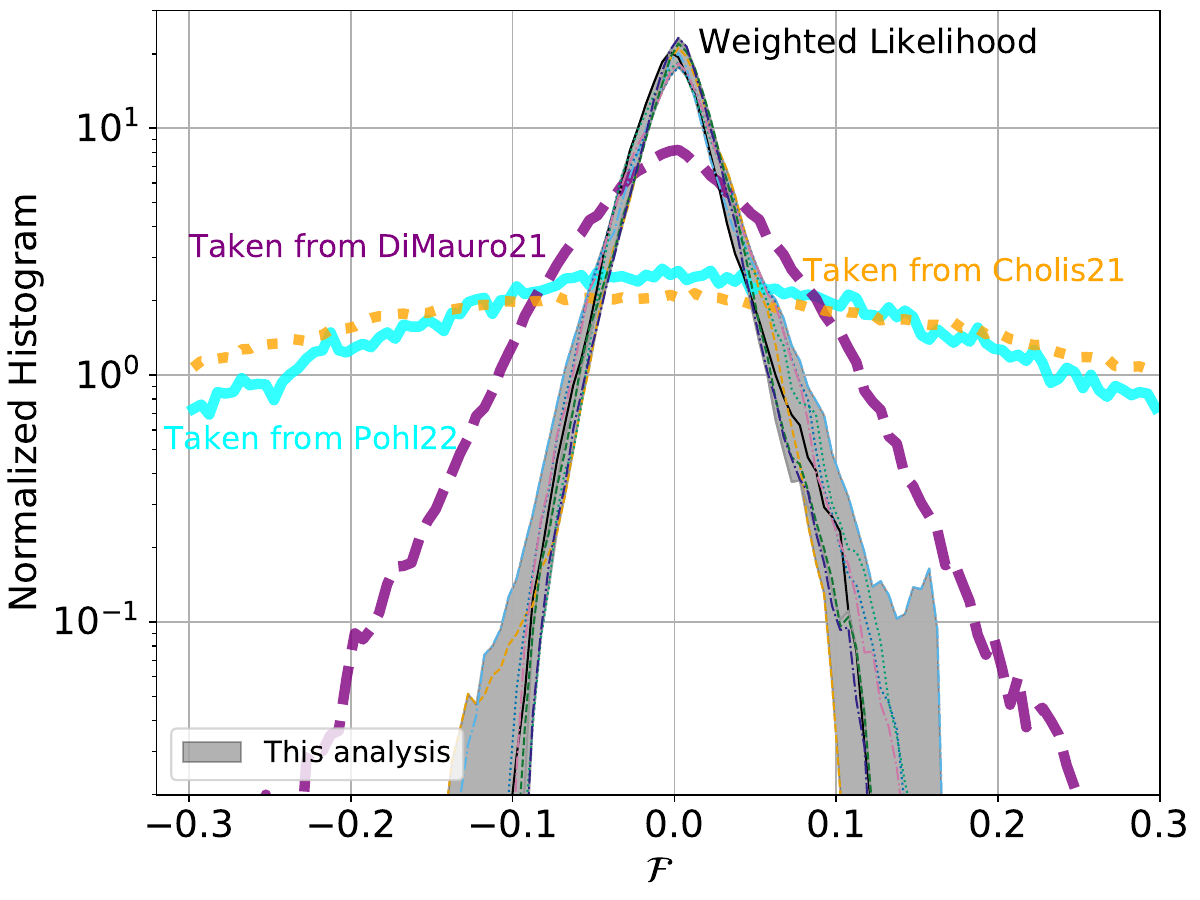}
  \includegraphics[width=0.49\linewidth]{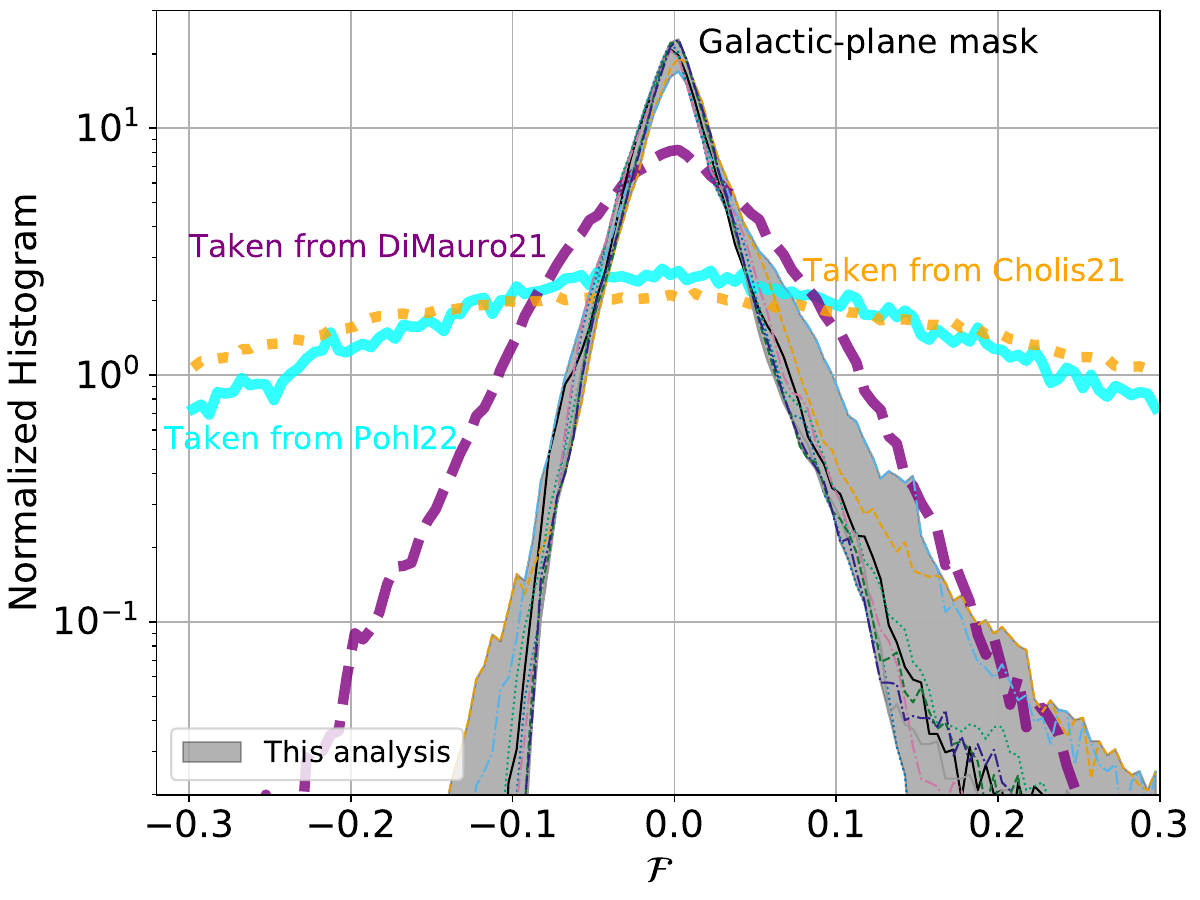}
  \caption{Normalized histograms of the fractional residuals $\mathcal{F}$ defined in Eq.~\ref{eq:Fapp}, shown for different analysis setups (panels). We overlay the residual distributions reported in \cite{DiMauro:2021prd} (DiMauro21, purple band), \cite{Pohl:2022nnd} (Pohl22, cyan band), and \cite{Cholis:2021rpp} (Cholis21, orange dotted). The colored curves and gray band show the outcome of our analysis for the suite of IEMs considered in this work.}
  \label{fig:residualsv}
\end{figure}

\begin{figure}
  \includegraphics[width=0.49\linewidth]{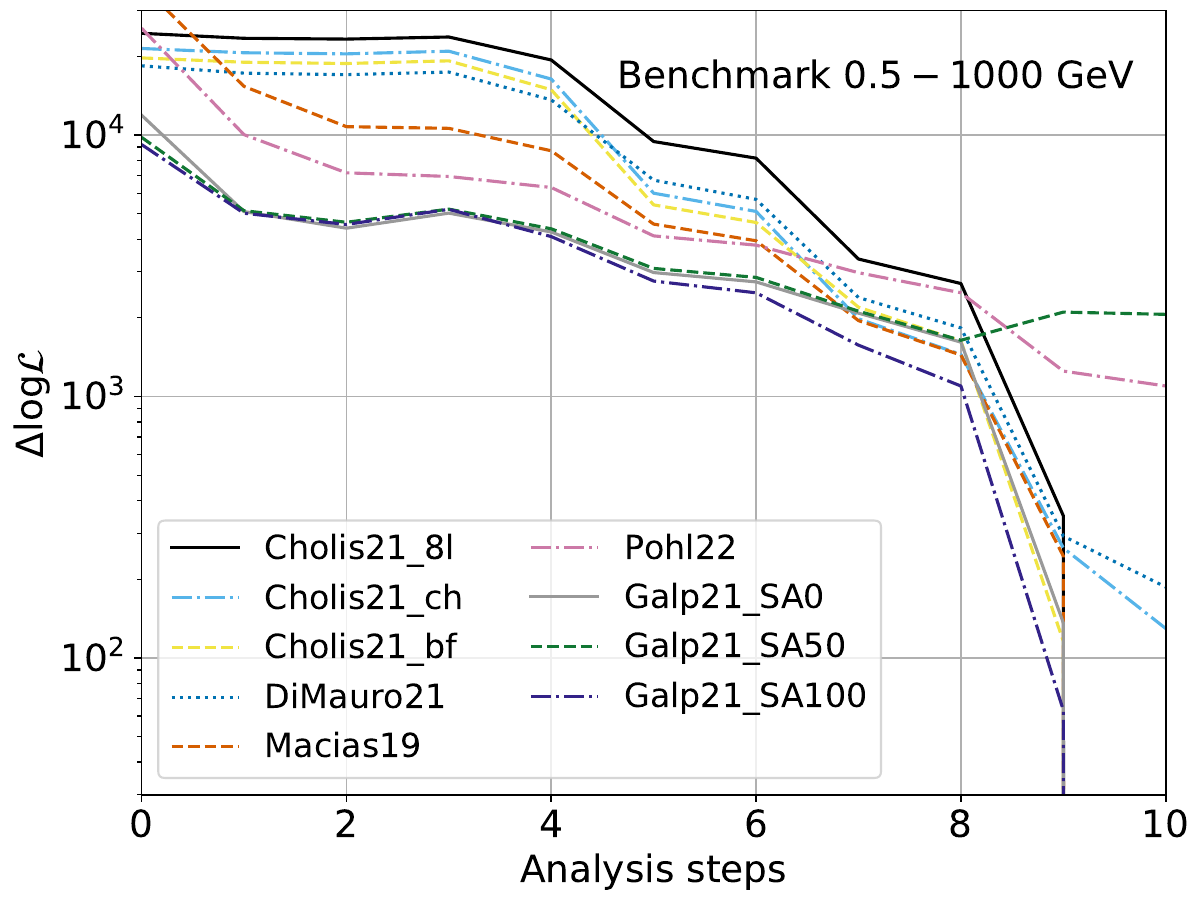}
  \includegraphics[width=0.49\linewidth]{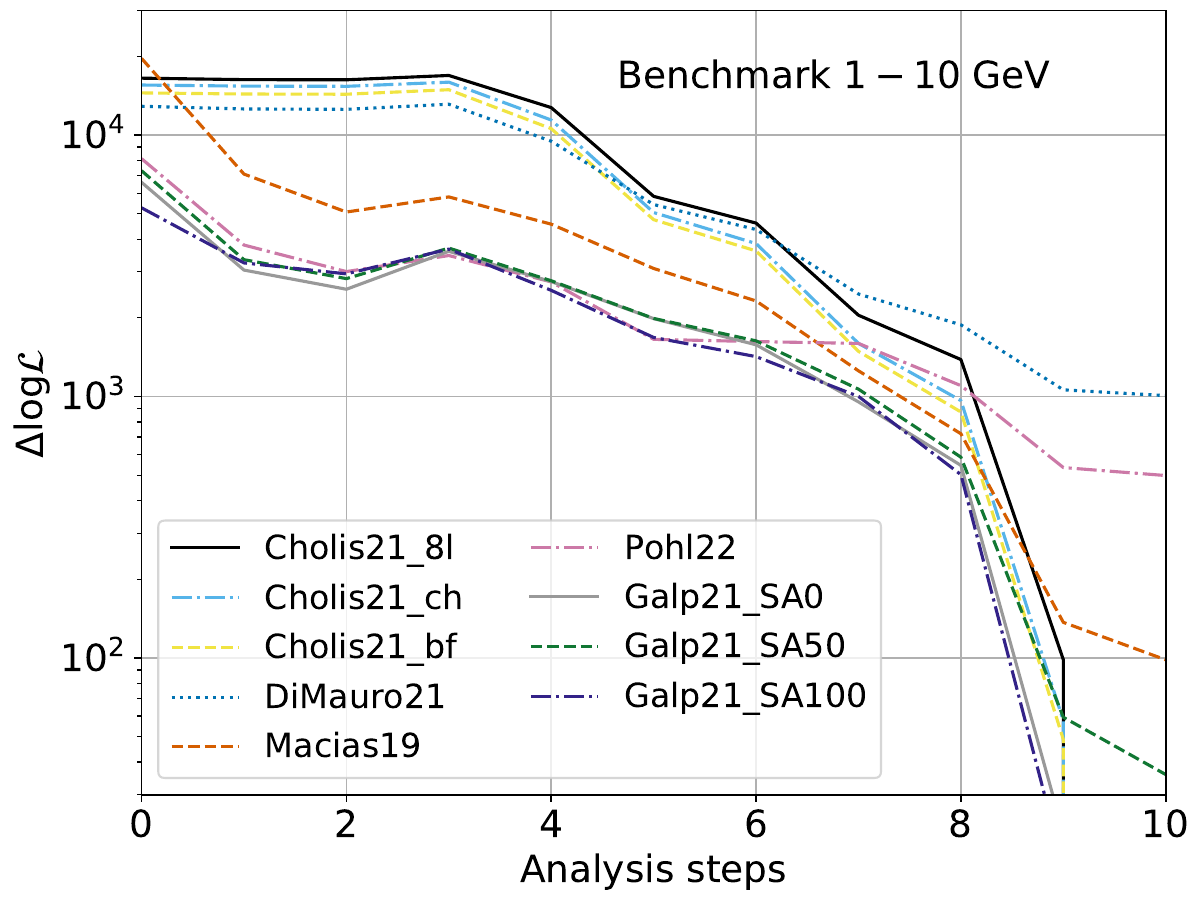}
  \includegraphics[width=0.49\linewidth]{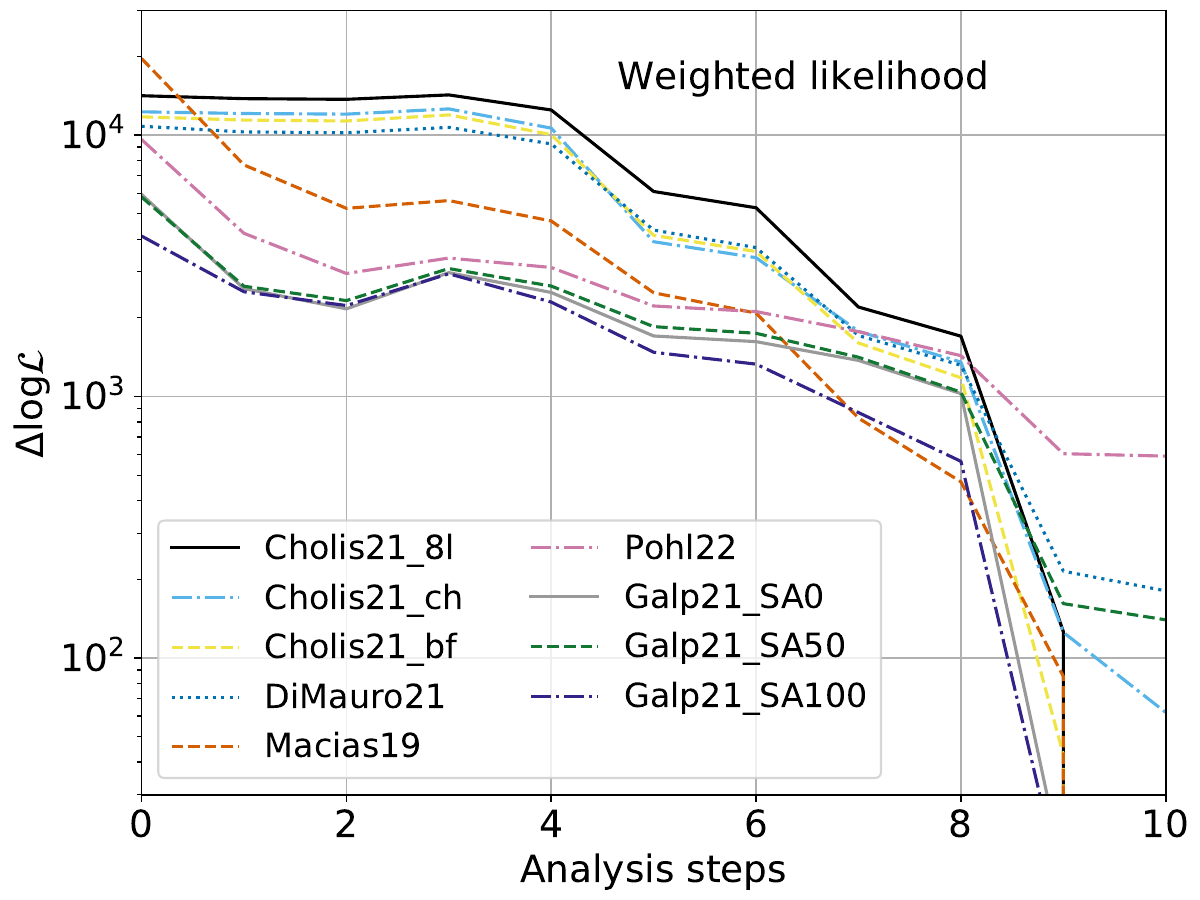}
  \includegraphics[width=0.49\linewidth]{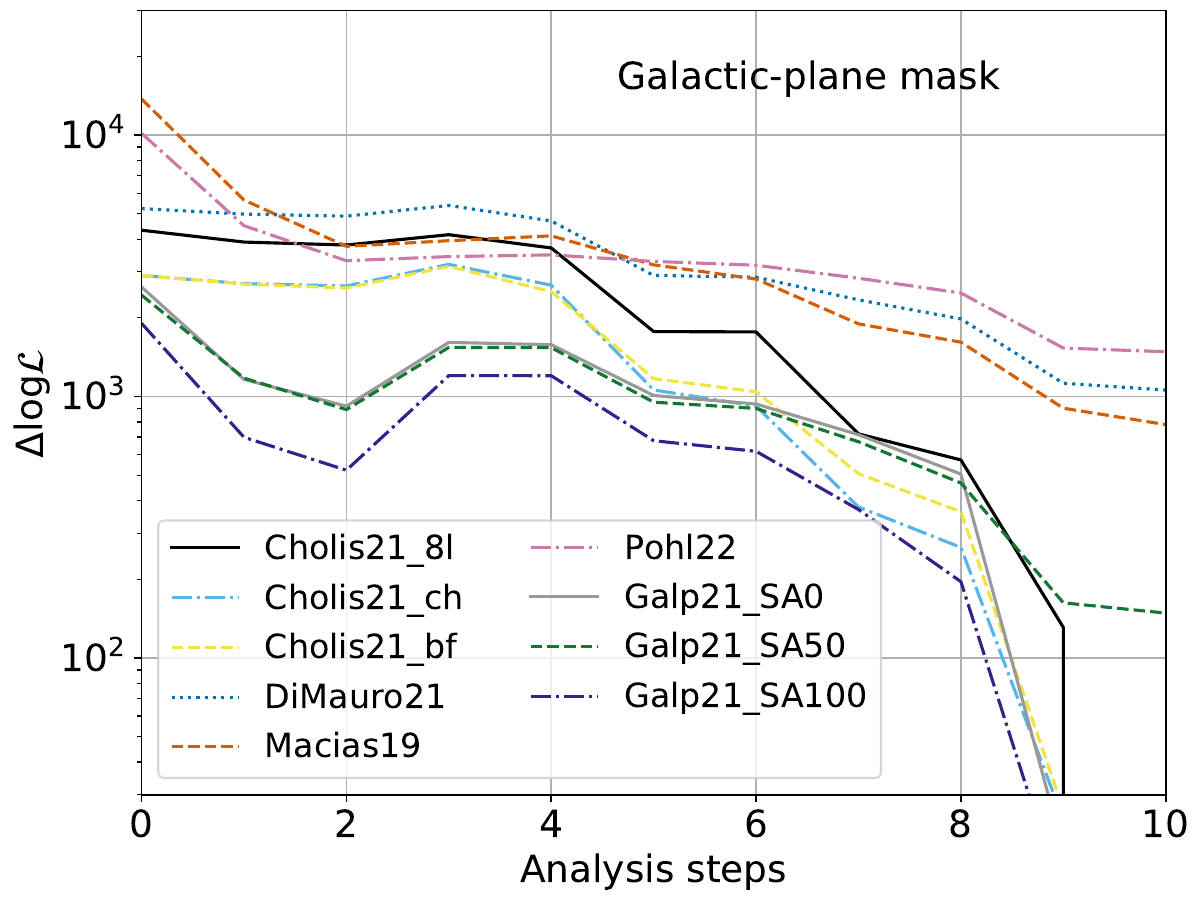}
  \caption{Evolution of $\Delta\log\mathcal{L}$ across the analysis steps. In each panel, $\Delta\log\mathcal{L}$ is shown relative to the best-fitting IEM at that step (i.e.\ the minimum among the IEMs in that panel). Different curves correspond to different IEMs. The top panels show the \textbf{Benchmark} setup (with the two energy ranges used in the main text), while the bottom panels show the \textbf{Weighted-Likelihood} (left) and \textbf{Galactic-plane mask} (right) setups.}
  \label{fig:like}
\end{figure}

Our analysis pipeline is implemented with \texttt{Fermipy}, a Python package providing a high-level interface to the \texttt{Fermitools} for standard binned-likelihood analyses of \emph{Fermi}-LAT data~\cite{2017ICRC...35..824W}.
We use \texttt{Fermipy} v1.2.0 together with \texttt{Fermitools} v2.2.0.

In our \textbf{Benchmark} analysis setup, we analyze a $40^\circ\times 40^\circ$ region of interest (ROI) centered on the Galactic Center (GC).
We use $\sim$16~years of \emph{Fermi}-LAT data from 2008-08-04 (MET=239557417) to 2024-11-28 (MET=754448190).
We select \texttt{P8R3\_SOURCEVETO} events and apply the standard data-quality selections (\texttt{DATA\_QUAL>0} and \texttt{LAT\_CONFIG==1}).
The data are binned using 8 energy bins per decade and a pixel size of $0.08^\circ$.
We apply energy dispersion to all sources and IEM templates using \texttt{edisp\_bins=-1}, i.e.\ including the adjacent energy bin in the energy-dispersion correction.
We select photons with energies $0.5$--$1000$~GeV when extracting the GCE SED and $1$--$10$~GeV for the morphological studies, to reduce diffuse-model systematics at low energies.
For $0.5$--$1$~GeV we require zenith angle $<100^\circ$, while for $E>1$~GeV we use $<105^\circ$, to minimize contamination from $\gamma$ rays originating from the Earth limb.
We include in the model all sources within a square of side $44^\circ$, i.e.\ extending 2$^\circ$ beyond the ROI boundary, to reduce edge effects from sources just outside the ROI.

This ROI contains $\mathcal{O}(500)$ cataloged sources, in addition to multiple extended and diffuse components.
A single global fit with all source parameters free is computationally prohibitive, both because the optimizer (based on \texttt{Minuit}) must explore a high-dimensional parameter space and because strong degeneracies among diffuse and extended components slow convergence.
We therefore adopt a staged fitting strategy designed to reach a stable likelihood maximum while controlling the dominant sources of mismodeling in the inner Galaxy:

\begin{itemize}
\item \textbf{Data preparation and initial optimization.}
We initialize the ROI and construct the binned data products with \texttt{gta.setup()}, applying the event selections in time and energy, the spatial and energy binning, and the chosen instrument response functions.
The main output of \texttt{gta.setup()} is the set of binned counts and exposure maps in Galactic longitude, latitude, and energy.
We then obtain an initial best-fit model using twice the tool \texttt{gta.optimize()}, which performs an iterative maximization of the likelihood by
(i) fitting the normalizations of the brightest diffuse components and sources simultaneously,
(ii) refitting the normalizations of the remaining sources, and
(iii) refining the solution by varying jointly source normalizations and spectral-shape parameters.
Here and in what follows when mentioning the brightest diffuse components and sources we refer in particular to case where their $TS>1000$.

\item \textbf{Full likelihood fit with a restricted free-parameter set.}
In regions with mild component degeneracies (e.g., high latitudes or analyses restricted to $E\gtrsim 1$~GeV), \texttt{gta.optimize()} typically converges to a solution close to that obtained from a single global likelihood maximization.
Toward the GC, however, intense diffuse emission, multiple extended templates, and the high source density produce strong correlations among parameters.
We therefore perform a full likelihood fit with \texttt{gta.fit()}, which wraps the \texttt{pyLikelihood} engine and returns best-fit parameters together with the covariance matrix.
To keep the computation tractable, we restrict the set of free parameters (normalizations and spectral slopes) by allowing the brightest diffuse components and the sources within $1^\circ$ of the GC to vary, while keeping the remaining sources fixed at their optimized values.

\item \textbf{Refining individual source spectra.}
We refit each source in the ROI with \texttt{gta.fit()}, while keeping free all the sources within $1^\circ$ from the source of interest and the brightest diffuse components.
After this step we rerun \texttt{gta.optimize()} and perform a subsequent fit including the DM (or bulge) templates and the diffuse components.

\item \textbf{Catalog pruning and refit.}
We prune the model by removing faint sources with $\mathrm{TS}<16$, consistent with an approximate $4\sigma$ inclusion threshold.\footnote{For nested hypotheses, $\sqrt{\mathrm{TS}}$ is often used as an approximate Gaussian significance, but this mapping is only valid under the conditions where Chernoff's theorem applies.}
After pruning, we rerun \texttt{gta.optimize()} and \texttt{gta.fit()} leaving free only the DM (or bulge), the IEM components and the brightest sources close to the GC region and obtain the likelihood solution used for subsequent steps.

\item \textbf{Iterative search for additional sources.}
We iteratively search for previously unmodeled sources using \texttt{gta.find\_sources}, in three passes with thresholds $\mathrm{TS}>100$, then $\mathrm{TS}>64$, and finally $\mathrm{TS}>36$.
For candidates above each threshold, we also test for spatial extension where appropriate.
Depending on the IEM, we identify $\sim 30$--$50$ additional source candidates, several with $\mathrm{TS}>100$ and extensions of order a degree.
When searching for spatial extension we place an upper limits of $2^\circ$. Most of these candidates cluster along the Galactic plane, where diffuse-model imperfections are known to be largest.
Including these sources improves the global likelihood and reduces the risk that structured mismodeling of the plane propagates into biased residuals at intermediate latitudes where the GCE contributes non-negligibly.
The 4FGL cataloged and newly detected sources are overlaid on the residual maps in Fig.~\ref{fig:resmap}.

\item \textbf{Final refit and bin-by-bin SED extraction.}
After adding the new sources, we rerun \texttt{gta.optimize()} and \texttt{gta.fit()}, leaving free only the DM (or bulge), the IEM components and the brightest sources closer than $5^{\circ}$ to the GC region, to obtain the final global best fit.
We then extract the GCE SED using \texttt{gta.sed}, which performs a bin-by-bin likelihood analysis by fitting an independent flux normalization in each energy bin while assuming a power law with fixed index $\Gamma=2.0$ within the bin.
Given the narrow binning adopted here, this approximation does not bias the recovered bin-by-bin fluxes.
In the bin-by-bin fits we leave free the normalizations of the GCE template, the brightest interstellar-emission components, and the sources with $\mathrm{TS}>10^4$ located within $2^\circ$ of the GC.
With this choice, the extracted GCE SED is effectively independent of the spectral model assumed for the GCE template in the global fit, enabling model-agnostic comparisons to specific DM or astrophysical scenarios \emph{a posteriori}.
\end{itemize}

We note that our multi-step pipeline introduces additional degrees of freedom mainly to capture known instrumental and astrophysical complexities of the inner Galaxy (diffuse components, extended templates, and previously unmodeled sources) and to reach a stable likelihood maximum in a high-dimensional parameter space, rather than to ``fine-tune'' the model to noise.
In practice, the main risk of over-flexibility would be an artificial absorption of the excess into background components; we test this explicitly by repeating the analysis with alternative IEMs and analysis setups that reduce model freedom (weighted-likelihood down-weighting of the brightest diffuse pixels and an explicit Galactic-plane mask), and we find that the recovered GCE spectrum and morphology remain stable within the quoted IEM-driven systematics.
Moreover, the GCE is extracted through a bin-by-bin SED procedure and through annular-profile fits in which the GCE normalizations are free in each bin/annulus, so the results are not tied to a highly tuned global spectral parameterization; the persistence of the signal across IEMs and analysis variants indicates that potential fine-tuning of nuisance parameters does not drive our conclusions.

As explained in the main text, we use different analysis setups.
The \textbf{Benchmark} setup uses the data selection described above and two energy ranges: $1$--$10$~GeV for morphological studies (to reduce low-energy systematics and because the IRFs are most favorable) and $0.5$--$1000$~GeV for the GCE SED.

Because the GC region is dominated by bright, structured diffuse emission, residual mismodeling of the Galactic plane can bias the fits and inflate test statistics, especially at low energies.
We therefore complement the \textbf{Benchmark} setup with two alternative analysis choices, both using \texttt{SOURCEVETO} events in the range $0.5$--$1000$~GeV.
First, we apply the \textbf{Weighted-Likelihood} method adopted in \emph{Fermi}-LAT analyses to mitigate diffuse-model systematics~\cite{Bruel:2021mrd,Fermi-LAT:2019yla}; each spatial/energy bin is assigned a weight $\epsilon\in[0,1]$ derived from the model-predicted foreground intensity, so that bins with large expected backgrounds are down-weighted and do not dominate the fit (typically $\epsilon\ll 1$ only within a few degrees of the Galactic plane and below $\sim$1~GeV).
This is illustrated in Fig.~\ref{fig:weightmap}, which shows the weight maps at $E=0.6$~GeV and $E=1.5$~GeV.
At $E=0.6$~GeV and near the Galactic plane, $\epsilon$ can be as small as $\sim 0.02$--$0.1$, whereas at $E=1.5$~GeV most of the ROI has $\epsilon\simeq 0.8$--$1.0$.
Thus, the weighted-likelihood approach mainly impacts the Galactic-plane region and energies below a few GeV.
Second, we repeat the analysis with a \textbf{Galactic-plane mask} $|b|<2^\circ$, which removes the brightest diffuse emission and the regions where gas-template imperfections are largest, at the cost of reduced statistics and diminished sensitivity to the GCE.

Fig.~\ref{fig:resmap} shows the ROI maps of the fractional residuals, defined as
\begin{equation}
\label{eq:Fapp}
\mathcal{F}_i = 2\cdot \frac{\mathcal{C}_{M,i}-\mathcal{C}_{D,i}}{\mathcal{C}_{M,i}+\mathcal{C}_{D,i}},
\end{equation}
where $\mathcal{C}_{M,i}$ ($\mathcal{C}_{D,i}$) is the number of model (data) counts in the $i$-th pixel, evaluated after the final refinement stage.
Across the representative IEMs shown, the residuals are largely balanced between positive and negative excursions and do not exhibit prominent large-scale coherent structures within the ROI.
The remaining mismodeling reaches at most $|\mathcal{F}_i|\sim 10$--$15\%$.
This behavior is further quantified by the histograms of $\mathcal{F}$ shown in Fig.~\ref{fig:residualsv}.
For the $0.5$--$1000$~GeV selections, we find that the vast majority of pixels have $|\mathcal{F}_i|\lesssim 10\%$, with only a small fraction contributing to the tails.
When restricting to $1$--$10$~GeV, the distribution broadens because the photon statistics are lower; correspondingly, small absolute count differences translate into larger fractional residuals, while the statistical significance of these differences is reduced due to larger Poisson uncertainties.
We also note that the $\mathcal{F}$ distributions obtained in our analysis are substantially more peaked around zero than those reported in previous works such as \cite{DiMauro:2021prd} ($E>300$~MeV), \cite{Pohl:2022nnd} ($E>600$~MeV), and \cite{Cholis:2021rpp} ($E>500$~MeV), for which residuals can reach the $30$--$40\%$ level.

Fig.~\ref{fig:like} summarizes the evolution of $\Delta\log\mathcal{L}$ across the successive stages of the pipeline for the different analysis setups.
Across IEMs, the largest improvements occur during the source-finding and refitting stages, with typical decreases in $\Delta\log\mathcal{L}$ of order a few $10^4$.
By construction, the final steps exhibit only modest additional changes, indicating that the procedure has reached a stable likelihood maximum.
The same figure also highlights the relative performance of the different IEM families.
In these configurations, the \texttt{Galp21} models yield the best overall likelihoods, whereas some of the \texttt{Cholis21} realizations are disfavored by $\Delta\log\mathcal{L}$ at the level of a few $10^2$ relative to the best-performing IEM.
Notably, the \texttt{Pohl22} model, which was argued in Ref.~\cite{Pohl:2022nnd} to provide an improved diffuse description, is among the poorer-performing models in our pipeline, with $\Delta\log\mathcal{L}$ reaching the $10^3$ level in some steps.

\section{Results across analysis setups and IEM choices}
\label{app:results}

Table~\ref{tab:E_summarys} summarizes the likelihood comparisons and TS values obtained when repeating the analysis with different event selections/analysis variants and with the full suite of IEMs.
Here we focus on how the main trends evolve across setups, complementing (and not repeating) the discussion in the main text.

\paragraph*{Global fit quality and IEM ranking.}
Across all analysis variants, the \texttt{Galp21} family consistently provides the best overall likelihood, with \texttt{Galp21\_SA100} typically yielding the largest $\log\mathcal{L}_{\rm DM}$ and therefore defining the reference for $\Delta\log\mathcal{L}_{\rm DM}$.
The spread in $\Delta\log\mathcal{L}_{\rm DM}$ increases when including lower-energy photons (\textbf{Benchmark $0.5-1000$ GeV}), reflecting the stronger sensitivity of the global fit to diffuse-model imperfections at low energies.
In that configuration, the \texttt{Cholis21}, {\tt DiMauro21} and {\tt Pohl22} perform with a worse goodness of fit with $\Delta\log{\mathcal{L}}$ of the order of $(1-3)\cdot 10^3$. 
The \textbf{Weighted-likelihood} setup reduces the relative impact of the brightest diffuse regions and correspondingly compresses the dispersion in $\Delta\log\mathcal{L}_{\rm DM}$ that reaches maximum values of about $1.7 \cdot 10^3$, while the \textbf{Galactic-plane mask} similarly stabilizes the fit results across the different IEMs by removing the most problematic sky region, at the cost of reduced statistics.

\paragraph*{Persistence of the DM-like component.}
Independently of the analysis setup, the DM-motivated template is detected with high significance for all IEMs, with TS$_{\rm DM}$ remaining at the level of $\mathcal{O}(10^4)$ (and reaching even larger values in configurations with higher photon statistics).
While the absolute TS values change across setups---as expected when modifying energy selections or down-weighting/masking the Galactic plane---the qualitative conclusion is unchanged: a DM-like excess component is always preferred over the no-excess hypothesis within each IEM realization.
The \texttt{Pohl22} model continues to yield systematically smaller TS$_{\rm DM}$ than other IEMs, indicating a stronger redistribution of emission among diffuse components in that particular construction.

\paragraph*{Bulge-only replacement versus joint fits.}
Replacing the DM template with NB+BB (column $\Delta\log\mathcal{L}_{\rm NB}$) worsens the fit in the \textbf{Benchmark}, \textbf{Galactic-plane mask}, and \textbf{Weighted-likelihood} configurations with the only exception of the \texttt{Pohl22} model illustrating another time that this IEM model presents a special case.
Importantly, however, in all setups the combined model that includes both DM and NB+BB yields an improvement relative to DM alone ($\Delta\log\mathcal{L}_{\rm DNB}>0$), but typically only at the level of a few tens to a few hundreds.
This behavior indicates that bulge-correlated structures can absorb part of the residual emission, yet they do not eliminate the preference for the DM-like component: TS$_{\rm DM}^{\rm DNB}$ remains large across all IEMs and analysis choices.

\paragraph*{NB without BB.}
The DM+NB-only variant (columns $\Delta\log\mathcal{L}_{\rm NDM}$ and TS$_{\rm N}^{\rm NDM}$) provides, in most cases, only a modest improvement over DM alone, with the NB template ranging from marginal to highly significant depending on the IEM and setup.
The largest NB significances occur for \texttt{Pohl22} and in the lower-energy configuration, consistent with stronger degeneracies (and therefore greater flexibility) when low-energy diffuse emission is included.
Small or even slightly negative fitted TS values for NB in a few cases should be interpreted as consistent with no preference for that component within numerical accuracy and the adopted fitting strategy.

Overall, Table~\ref{tab:E_summarys} shows that the main conclusions of the Letter are robust: the DM-like excess persists with high significance across IEMs and analysis setups, bulge templates can improve the fit modestly but do not remove the DM preference and significance, and the relative partition between DM-like and bulge-like morphologies remains IEM dependent, motivating the systematic treatment of diffuse-model uncertainties adopted throughout this work.

\begin{figure}
  \centering
  \includegraphics[width=0.49\linewidth]{SB_GCE_40X40_110GeV_cleaned.pdf}
  \includegraphics[width=0.49\linewidth]{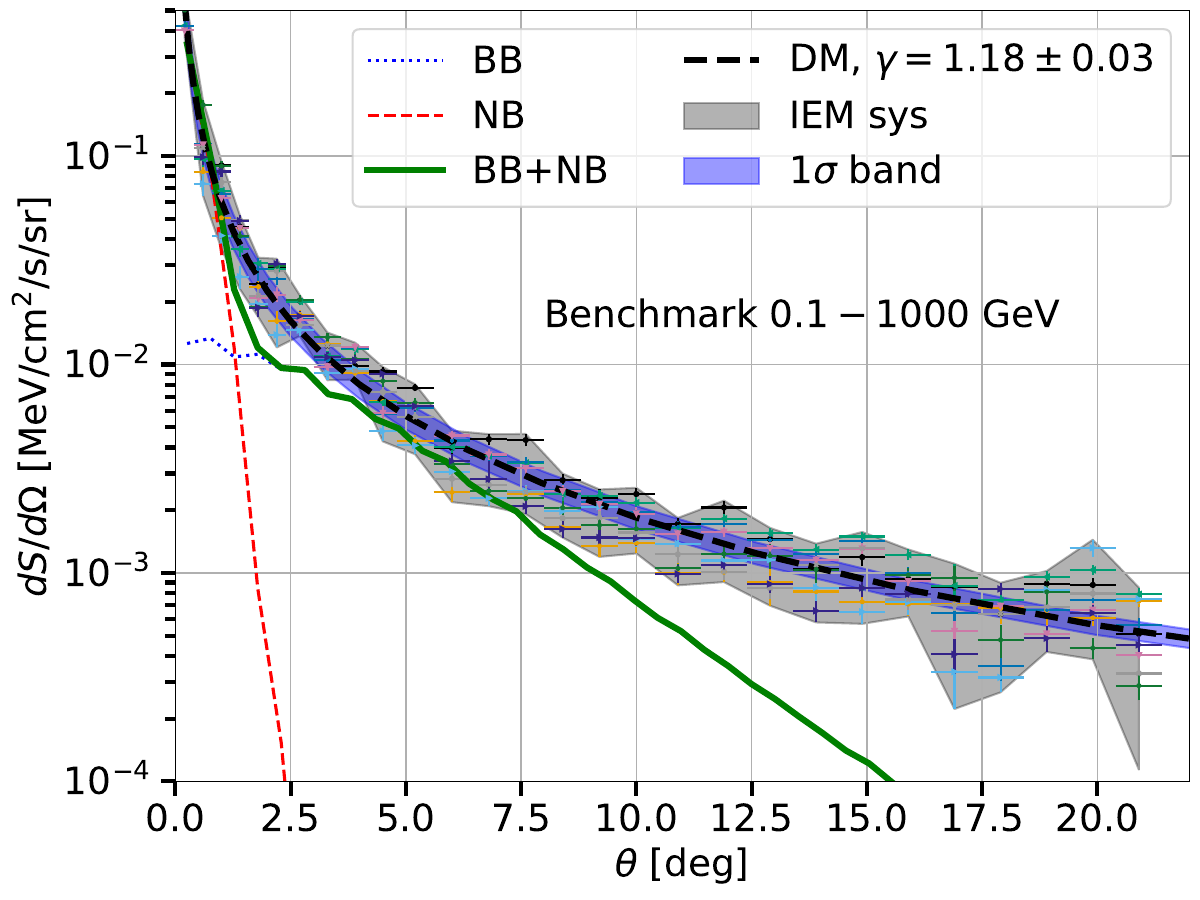}
  \includegraphics[width=0.49\linewidth]{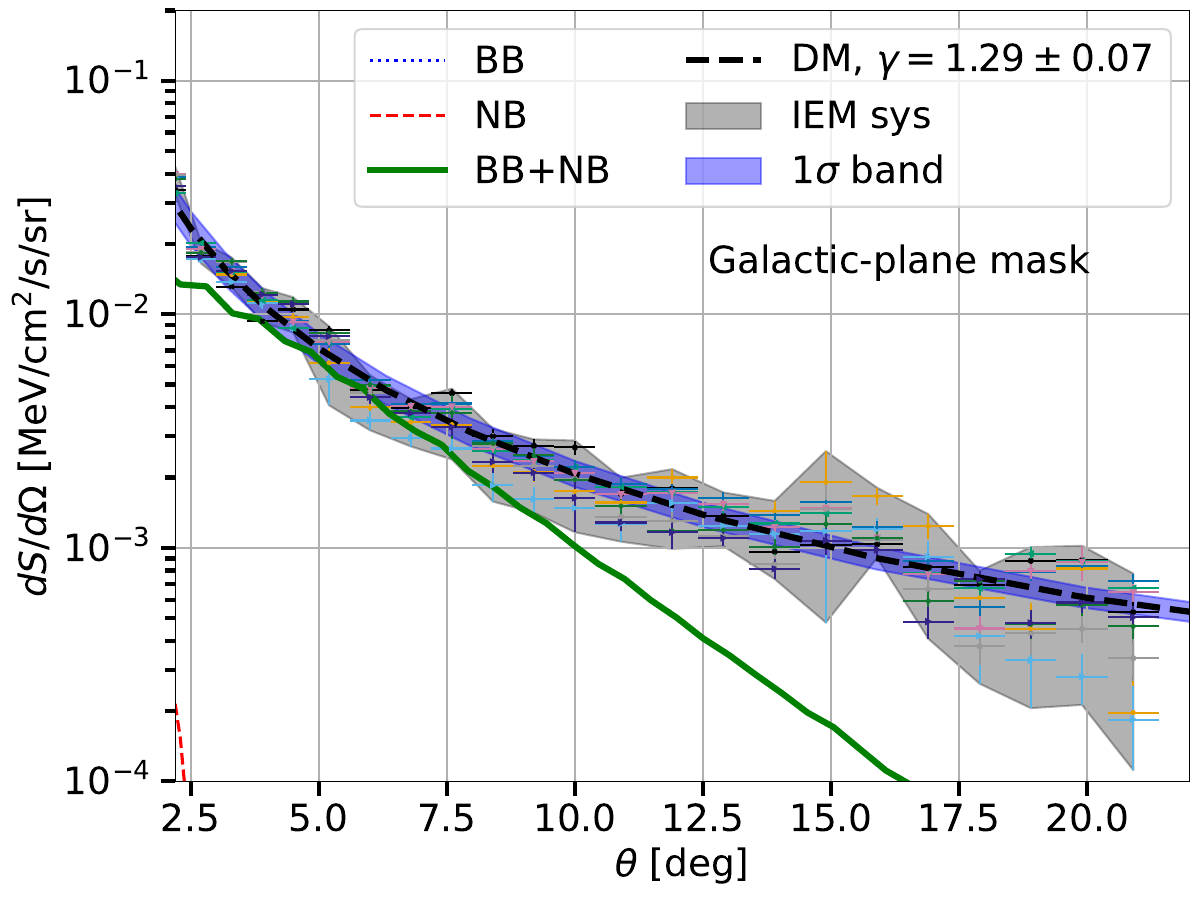}
  \includegraphics[width=0.49\linewidth]{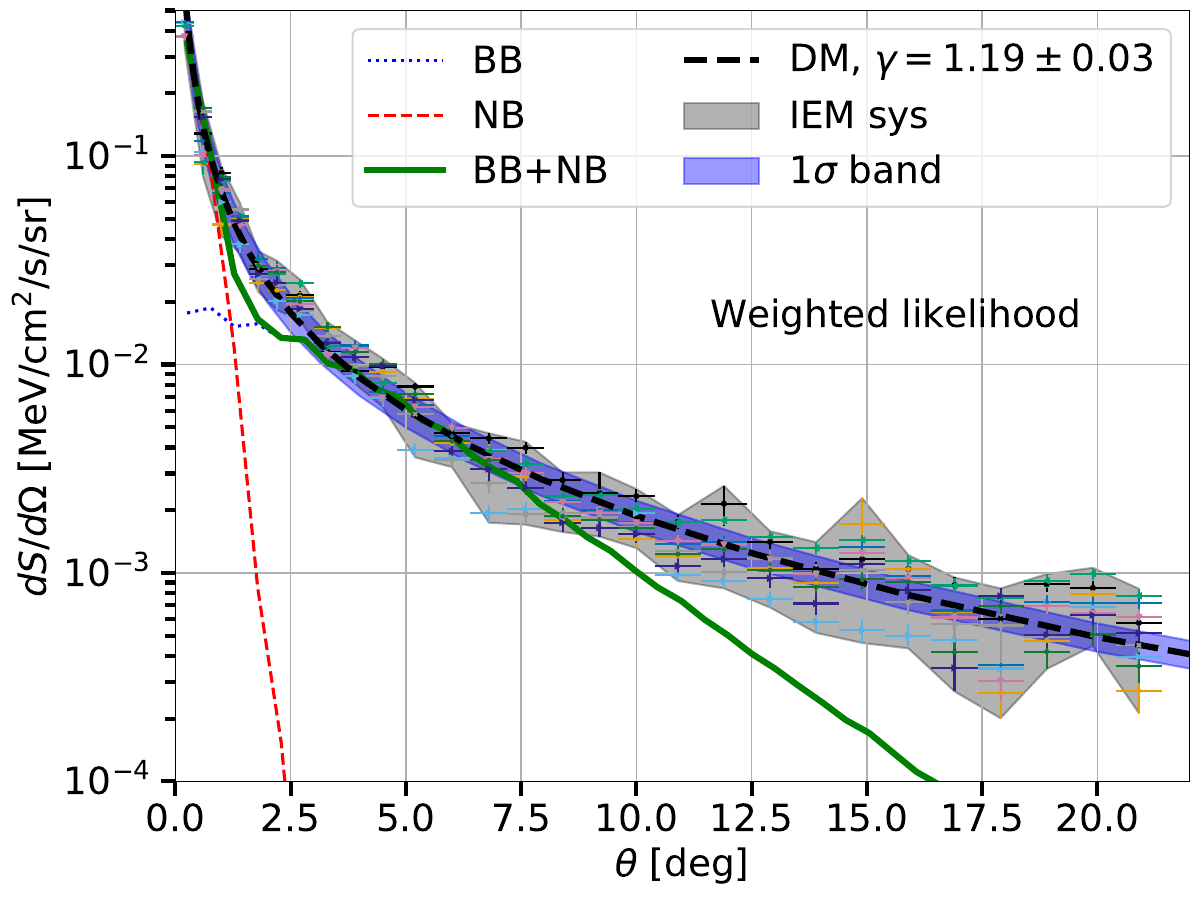}
  \caption{Surface brightness, $dS/d\Omega$, as a function of angular distance from the GC line of sight, reconstructed using the set of IEMs adopted in this work (colored points). The black dashed curve shows the best-fit gNFW profile ($\gamma=1.15$) with its $1\sigma$ uncertainty band (blue). The gray band indicates the spread associated with the IEM choice. The red dashed and blue dotted curves show the NB and BB components obtained from a fit using only NB+BB templates, with their sum shown in green. We show the results obtained for the four analysis setups used in the paper.}
  \label{fig:SBapp}
\end{figure}

\begin{figure}
  \centering
  \includegraphics[width=0.49\linewidth]{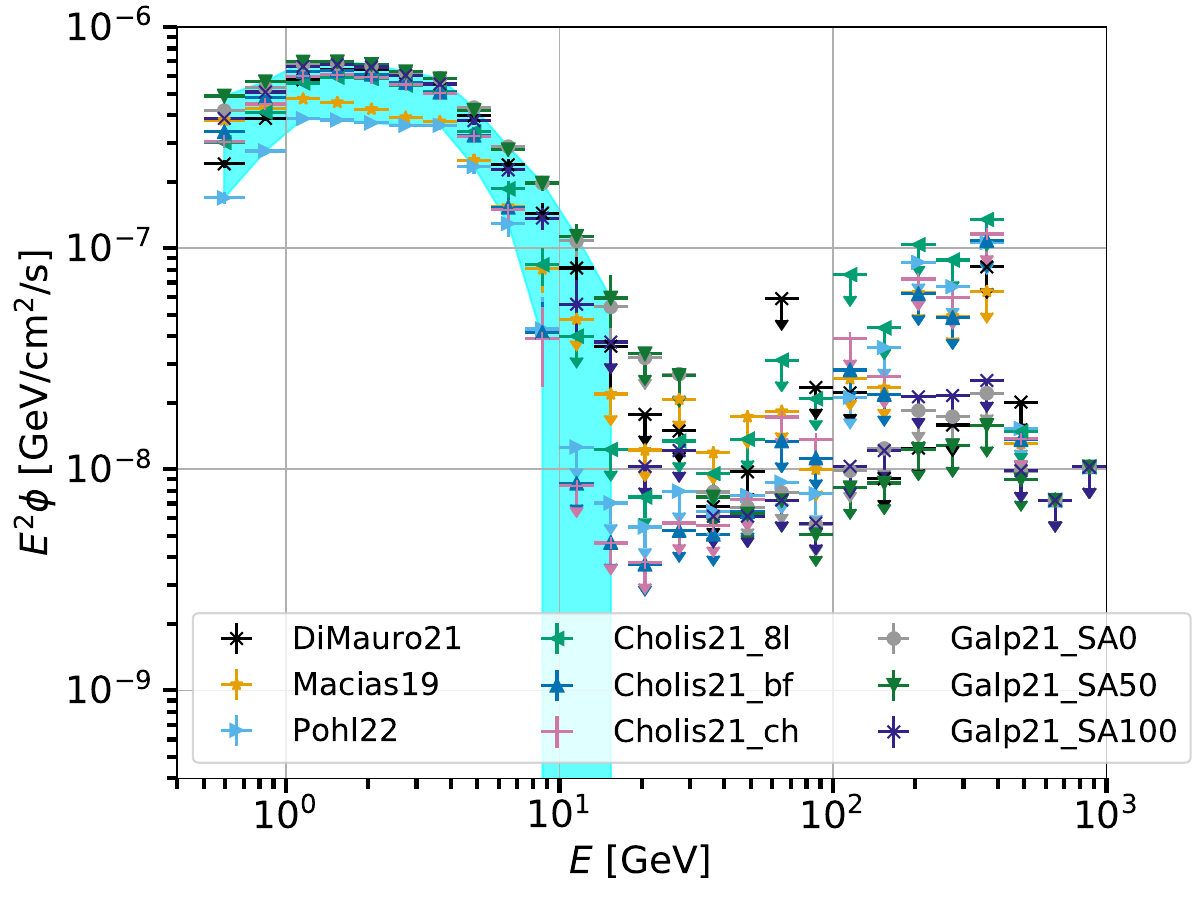}
  \caption{GCE SED obtained with our analysis and the different IEM models. We also display the envelope of all the spectra obtained with the different IEMs.}
  \label{fig:SEDdata}
\end{figure}

\begin{figure}
  \centering
  \includegraphics[width=0.49\linewidth]{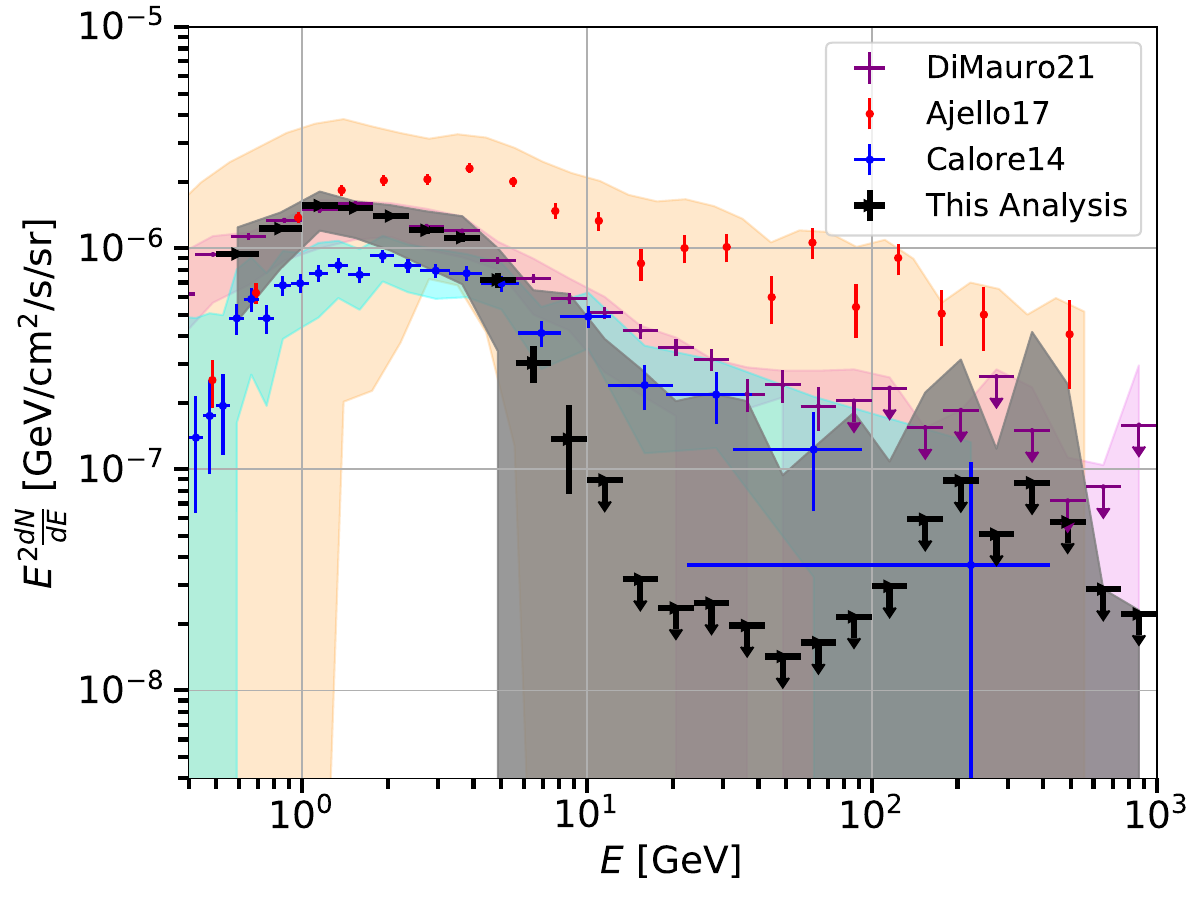}
  \includegraphics[width=0.49\linewidth]{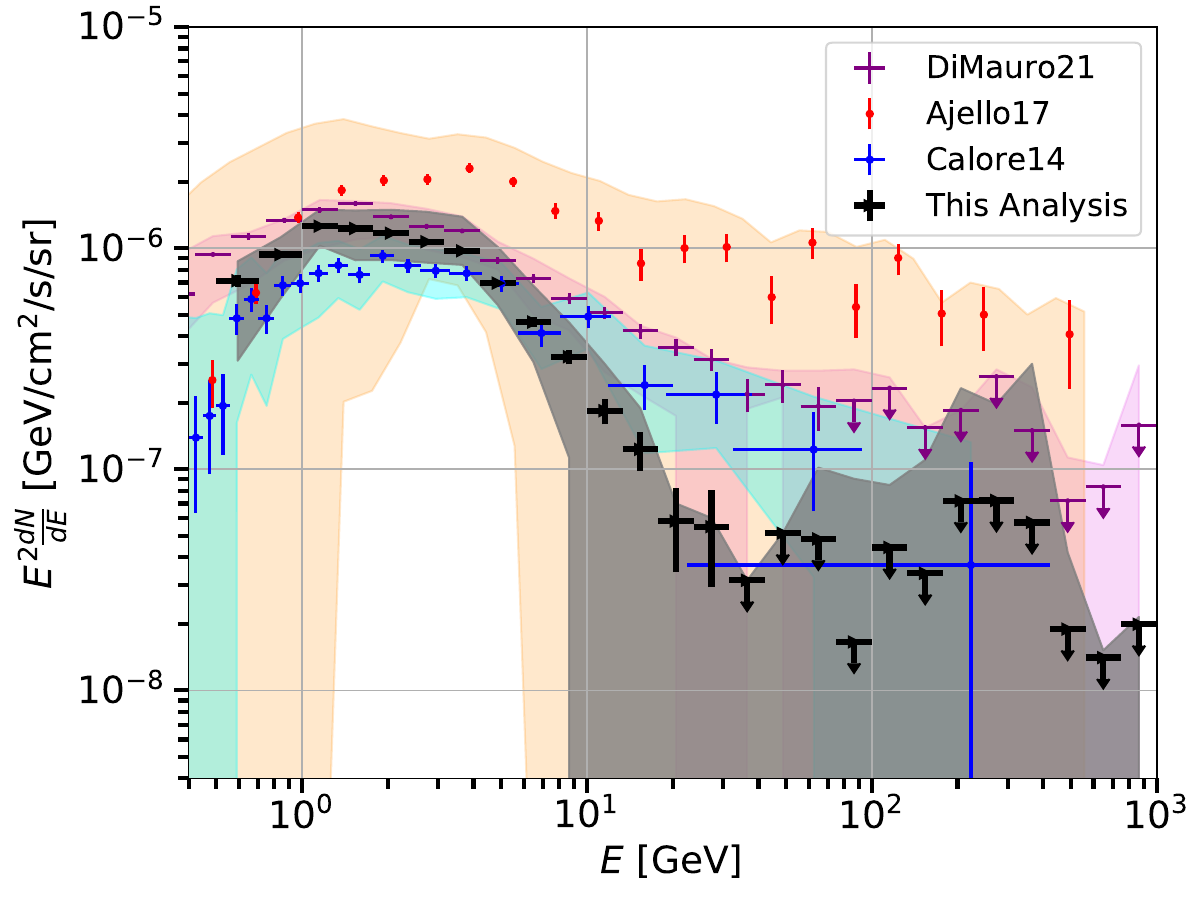}
  \caption{GCE SED obtained with our analysis compared with previous papers. In particular, we display our result obtained with the {\tt Galp21\_SA100} and the envelope of all the spectra obtained with the different IEMs. We also show the flux reported in Refs.~\cite{Calore:2014nla,Fermi-LAT:2017opo,DiMauro:2021prd}. We display the results obtained with the {\tt Galactic-plane mask} (left panel) and {\tt weighted-likelihood} (right panel) analysis setups.}
  \label{fig:SEDcompare}
\end{figure}

Figure~\ref{fig:SBapp} compares the reconstructed surface brightness for the four analysis configurations used in this work (two \textbf{Benchmark} energy selections, the \textbf{Galactic-plane mask}, and the \textbf{Weighted-likelihood} setup). 
In all cases, and for all IEMs (colored points), the profile is strongly centrally concentrated and is well described by a gNFW morphology over the full angular range shown. 
The best-fit inner slope $\gamma$ is stable with values between $1.1-1.2$ between the two \textbf{Benchmark} selections and the \textbf{Weighted-likelihood} analysis (see legends): the recovered profiles overlap within the quoted $1\sigma$ bands and the IEM-induced envelope (gray band), indicating that the extracted morphology at intermediate latitudes is not driven by the brightest Galactic-plane regions and is robust against the specific treatment of low-energy diffuse mismodeling.
The \textbf{Galactic-plane mask} setup yields a somewhat steeper preferred slope and a larger uncertainty. 
This behavior is expected because masking $|b|<2^\circ$ removes a substantial fraction of the photons and suppresses the region where the template degeneracies are strongest, thereby changing the leverage on the innermost annuli and increasing the statistical (and effective systematic) uncertainty of the reconstructed profile. 
Importantly, even in this conservative configuration the data points remain incompatible with a pure NB+BB interpretation: the NB component drops too rapidly beyond the inner degree and the summed NB+BB profile undershoots the reconstructed surface brightness at large angles, whereas a gNFW-like component continues to provide a good description.

We have also tested the impact of using different {\it Fermi}-bubble templates in the surface brightness. In particular, the template of the model {\tt DiMauro21} was taken from \cite{Fermi-LAT:2017opo}, which is divided into and inner galaxy and out galaxy components that are fitted individually to the data.
For the {\tt Galp21} models we use the same bubble model as in {\tt DiMauro21}. 
Instead, {\tt Cholis21} models the author use the more outdated results taken from \cite{2010ApJ...724.1044S}.
Finally, the IEM setups {\tt Pohl22} and {\tt Macias19}  take into account the templates in \cite{Fermi-LAT:2014sfa}.
In order to test the effect of different Bubble templates in the surface brightness, we apply to the Bubble model {\tt DiMauro21} and {\tt Galp21} the template used in {\tt Macias19} and vice-versa.
The results obtained are within the error consistent with the one shown in Fig.~\ref{fig:SBapp}.
Therefore, the choice of the {\it Fermi}-bubble templates does not affect our results.

Figure~\ref{fig:SEDdata} illustrates how the reconstructed GCE SED varies across the different IEM choices within the \textbf{Benchmark $0.5$--$1000$~GeV} setup.
While the overall spectral shape is stable---a peak at a few GeV followed by a rapid decline---the IEM choice mainly affects the normalization and the size of the error bars, particularly in the transition region around $\sim 10$--$30$~GeV where the extraction becomes more sensitive to diffuse-model degeneracies.
We therefore use the envelope spanned by the individual IEM realizations as a practical estimate of the residual diffuse-model systematic uncertainty on the SED.
At higher energies the spectra become increasingly background limited and are dominated by upper limits; in this regime, the spread among IEMs is consistent with the absence of a significant hard component and provides a robust bound on any additional contribution above $\sim 10$~GeV, such as the IC emission expected in bulge-MSP scenarios. In particular, at around $20-50$ GeV the upper limits obtained for the different IEM are all at the level of $E^2\Phi \lesssim (1-2)\cdot 10^{-8}$~GeV\,cm$^{-2}$\,s$^{-1}$\,sr$^{-1}$.
This outcome remains stable against the different analysis setup used in the paper.

Figure~\ref{fig:SEDcompare} shows the GCE SED obtained with the two additional analysis variants discussed in the text, namely the \textbf{Galactic-plane mask} (left panel) and the \textbf{Weighted-likelihood} setup (right panel), and compares them to representative determinations in the literature.
In both cases, the reconstructed spectrum is consistent with our \textbf{Benchmark} result within the IEM-driven envelope: the peak at a few GeV is preserved, and the emission rapidly decreases above $\sim$10--$30~\mathrm{GeV}$, with only upper limits at higher energies.
This demonstrates that the main spectral features---and in particular the absence of a pronounced hard component at tens of GeV and above---are not driven by the treatment of the Galactic plane, but remain stable when either masking the plane or down-weighting the brightest diffuse regions.

\begin{table}[t]
\centering
\begin{tabular}{lrrrrrrrrrr}
\hline
Model & $\Delta\log\mathcal{L}_{\rm DM}$ & TS$_{\rm DM}$ & $\Delta\log\mathcal{L}_{\rm NB}$ & TS$_{\rm B/N}$ & $\Delta\log\mathcal{L}_{\rm DNB}$ & TS$_{\rm DM}^{\rm DNB}$ & TS$_{\rm B/N}^{\rm DNB}$ & $\Delta\log\mathcal{L}_{\rm NDM}$ & TS$_{\rm DM}^{\rm NDM}$ & TS$_{\rm N}^{\rm NDM}$ \\
\hline

\multicolumn{11}{c}{\textbf{Benchmark setup $1-10$ GeV}}\\
\hline
Cholis21\_8l & -2148 & 12772 & -513 & 4143/3677 & 18 & 11749 & 15/124 & 15 & 11997 & 123 \\
Cholis21\_ch & -1664 & 14091 & -500 & 5400/4203 & 82 & 8850 & 632/328 & 8 & 13691 & 51 \\
Cholis21\_bf & -1562 & 15141 & -681 & 5223/4189 & 53 & 10871 & 373/239 & 8 & 14757 & 52 \\
DiMauro21 & -2649 & 17198 & -566 & 4024/10369 & 48 & 15259 & 2/566 & 48 & 15338 & 563 \\
Macias19 & -1718 & 13231 & -109 & 6248/5352 & 132 & 7179 & 1769/625 & 10 & 12485 & 101 \\
Pohl22 & -2195 & 6137 & 343 & 6030/5519 & 347 & 764 & 3952/3704 & 121 & 4432 & 2763 \\
Galp21\_SA0 & -918 & 19875 & -131 & 9271/8207 & 201 & 8535 & 1988/1392 & 31 & 16889 & 429 \\
Galp21\_SA50 & -937 & 22909 & -256 & 9634/8770 & 194 & 10104 & 1639/1349 & 37 & 21073 & 353 \\
Galp21\_SA100 & 0 & 19294 & -242 & 8135/9733 & 124 & 9928 & 1150/2194 & 30 & 17714 & 718 \\

\hline
\multicolumn{11}{c}{\textbf{Benchmark setup $0.5-100$ GeV}}\\
\hline
Cholis21\_8l & -3041 & 22219 & -1025 & 7114/4824 & 59 & 21041 & 180/92 & 4 & 22359 & 1 \\
Cholis21\_ch & -3308 & 24024 & -676 & 10148/5592 & 155 & 15532 & 1037/756 & 42 & 22663 & 274 \\
Cholis21\_bf & -3098 & 25660 & -817 & 9555/6137 & 144 & 17187 & 830/880 & 58 & 24074 & 387 \\
DiMauro21 & -3315 & 25995 & -775 & 5213/14394 & 56 & 22143 & 33/816 & 49 & 22774 & 748 \\
Macias19 & -1383 & 23210 & -232 & 7179/9035 & 162 & 15822 & 1556/1104 & 39 & 21143 & 488 \\
Pohl22 & -2582 & 12865 & 474 & 9377/11779 & 482 & 1789 & 5925/5666 & 117 & 9261 & 3475 \\
Galp21\_SA0 & -650 & 27991 & -9 & 14419/16730 & 436 & 11291 & 3725/5957 & 146 & 22109 & 3962 \\
Galp21\_SA50 & -2727 & 33246 & 207 & 12436/16982 & 670 & 11779 & 2993/6354 & 290 & 25983 & 3768 \\
Galp21\_SA100 & 0 & 28937 & -237 & 8627/17707 & 272 & 12077 & 1574/6809 & 159 & 23397 & 4182 \\

\hline
\multicolumn{11}{c}{\textbf{Galactic-plane mask setup}}\\
\hline
Cholis21\_8l & -932 & 34320 & -817 & 8352/4786 & 27 & 32892 & 113/44 & 4 & 33728 & 47 \\
Cholis21\_ch & -950 & 31501 & -509 & 9524/2464 & 30 & 24406 & 254/234 & 14 & 30625 & 141 \\
Cholis21\_bf & -735 & 34362 & -565 & 9927/3209 & 41 & 27362 & 213/326 & 23 & 32951 & 243 \\
DiMauro21 & -1763 & 25477 & -555 & 4331/4009 & 40 & 24355 & 57/1077 & 34 & 25258 & 977 \\
Macias19 & -1513 & 52077 & -231 & 7178/6401 & 118 & 31664 & 1089/1334 & 62 & 41104 & 933 \\
Pohl22 & -2224 & 17025 & 240 & 8454/4858 & 267 & 2342 & 4953/2917 & 72 & 13325 & 1242 \\
Galp21\_SA0 & -283 & 40862 & -361 & 8520/3783 & 107 & 26213 & 974/281 & 14 & 38397 & 60 \\
Galp21\_SA50 & -430 & 41898 & -311 & 9433/4168 & 167 & 23145 & 1515/697 & 18 & 38473 & 259 \\
Galp21\_SA100 & 0 & 35376 & -313 & 12421/3365 & 119 & 22445 & 1552/327 & 1 & 32859 & 70 \\

\hline
\multicolumn{11}{c}{\textbf{Weighted-likelihood analysis setup}}\\
\hline
Cholis21\_8l & -1580 & 16494 & -891 & 6335/2559 & 51 & 15611 & 173/50 & 10 & 16108 & 40 \\
Cholis21\_ch & -1708 & 15392 & -408 & 7175/2351 & 99 & 9618 & 1006/238 & 9 & 15021 & 26 \\
Cholis21\_bf & -1534 & 16857 & -507 & 7316/2623 & 84 & 11217 & 818/238 & 11 & 16422 & 37 \\
DiMauro21 & -1721 & 17125 & -674 & 3973/6132 & 38 & 14839 & 57/401 & 27 & 15470 & 354 \\
Macias19 & -1041 & 13262 & -115 & 4914/3803 & 128 & 8338 & 1490/680 & 27 & 12403 & 160 \\
Pohl22 & -1708 & 8181 & 354 & 6687/4768 & 402 & 733 & 5872/3317 & 56 & 6689 & 729 \\
Galp21\_SA0 & -485 & 17588 & -28 & 10458/7142 & 293 & 6572 & 3859/2013 & 81 & 13628 & 830 \\
Galp21\_SA50 & -619 & 19198 & -55 & 7070/7384 & 261 & 7426 & 2247/1992 & 67 & 15651 & 691 \\
Galp21\_SA100 & 0 & 18208 & -144 & 5824/6789 & 185 & 7918 & 1194/1625 & 81 & 15076 & 697 \\

\hline
\end{tabular}
\caption{Summary of fit quality and test statistics for the different IEMs and analysis setups. 
Column~(2) reports $\Delta\log\mathcal{L}_{\rm DM}\equiv \log\mathcal{L}_{\rm DM}^{\max}-\log\mathcal{L}_{\rm DM}$, where $\log\mathcal{L}_{\rm DM}^{\max}$ is the largest DM-only log-likelihood within the corresponding setup (typically \texttt{Galp21\_SA100}). 
Column~(3) gives TS$_{\rm DM}$ for the DM-only fit. 
Columns~(4) and (5) report the change in log-likelihood and the TS values when replacing DM with NB+BB (TS$_{\rm B/N}$ is reported as BB/NB). 
Columns~(6)--(8) give $\Delta\log\mathcal{L}_{\rm DNB}$ and the corresponding TS values when adding NB+BB on top of DM (TS$_{\rm B/N}^{\rm DNB}$ is reported as BB/NB). 
Columns~(9)--(11) report $\Delta\log\mathcal{L}_{\rm NDM}$ and the TS values for the fit including DM+NB only (TS$_{\rm N}^{\rm NDM}$ refers to the NB component). 
Negative $\Delta\log\mathcal{L}$ values indicate a worse fit relative to the stated reference model. The different vertical block represent the results obtained with the different analysis setups.}
\label{tab:E_summarys}
\end{table}


\bibliographystyle{apsrev4-2}
\bibliography{main}

\end{document}